\documentclass{egpublarxiv}
\usepackage{cgfpreprint}

\JournalPaper         %

\CGFStandardLicense

\usepackage[T1]{fontenc}
\usepackage{dfadobe}  

\usepackage{cite}  %
\BibtexOrBiblatex
\electronicVersion
\PrintedOrElectronic
\ifpdf \usepackage[pdftex]{graphicx} \pdfcompresslevel=9
\else \usepackage[dvips]{graphicx} \fi

\usepackage{egweblnk} 
\usepackage[ruled]{algorithm2e} %

\SetAlFnt{\small}
\SetAlCapFnt{\small}
\SetAlCapNameFnt{\small}
\SetAlCapHSkip{0pt}

\usepackage{amsfonts}
\usepackage{amsmath}
\usepackage{subcaption}
\usepackage{multirow}
\usepackage{etoolbox}
\usepackage{amsbsy}
\usepackage{wrapfig}
\usepackage{algpseudocode}
\usepackage{float}
\usepackage{xcolor}

\usepackage[table]{xcolor}
\definecolor{best}{HTML}{8C8CFF}   %
\definecolor{second}{HTML}{B2B2FF} %

\DeclareMathOperator*{\argmax}{arg\,max}
\DeclareMathOperator*{\argmin}{arg\,min}

\usepackage[normalem]{ulem}

\newcommand{\R}{\mathbb{R}}
\newcommand{\T}{\mathbb{T}}
\newcommand{\etal}{\textit{et al.}}
\newcommand{\tableref}[1]{Tab.~\ref{#1}}
\newcommand{\figureref}[1]{Fig.~\ref{#1}}
\newcommand{\equationref}[1]{Eq.~\ref{#1}}
\newcommand{\equationsref}[1]{Eqs.~\ref{#1}}
\newcommand{\algoref}[1]{Alg.~\ref{#1}}
\newcommand{\sectionref}[1]{Sec.~\ref{#1}}
\newcommand{\MVStex}[0]{MVS-tex}
\newcommand{\ColorMap}[0]{ColorMap}
\newcommand{\AdvInitAlign}[0]{AdvIA}
\newcommand{\ImSurfTex}[0]{Im2SurfTex}
\newcommand{\appendixref}[1]{App.~\ref{#1}}
\usepackage{booktabs}

\newcommand{\Tighten}[1]{}
\newcommand{\citeauth}[2]{#1~\etal\cite{#2}}

\usepackage[ruled]{algorithm2e} %

\SetAlFnt{\small}
\SetAlCapFnt{\small}
\SetAlCapNameFnt{\small}
\SetAlCapHSkip{0pt}

\newcommand{\sneakInCaption}[2]{%
    \begin{subfigure}[b]{#1\textwidth}
        \centering
        \includegraphics[width=\textwidth,height=1px,keepaspectratio=false]{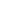}
        \centering
        \caption*{#2}
    \end{subfigure}
}

\def\Surface{\mathcal{M}}
\def\Render{Y}
\def\Camera{C}
\def\Color{I}
\def\Confidence{w}
\def\LocalColorMean{\mu}
\def\LocalColorScale{\sigma}
\def\NormalizedColor{\Color'}
\def\AdvectedColor{\hat{\Color}}
\def\AdvectedNormalizedColor{\hat{\Color}'}
\def\AdvectedConfidence{\hat{\Confidence}}
\def\SmoothedAdvectedNormalizedColor{\overline{\Color}'}
\def\FlowField{\vec{v}}
\def\NormalizationWeight{\epsilon_\text{n}}
\def\ScalarSmoothingWeight{\beta_\text{blur}}
\def\FlowFieldSmoothingWeight{\beta_\text{reg}}
\def\GradientFittingWeight{\alpha_\text{gf}}
\def\GradientDomainRegularizationWeight{\alpha_\text{reg}}
\def\OutputColor{I^*}
\def\ScalarBasis{\phi}
\def\VectorBasis{\vec{\nu}}
\def\SurfaceIntegral{\int_\Surface}

\newcommand{\revise}[1]{#1}

\title[Texture Consolidation of Generated Multi-View Renderings]%
      {Texture Consolidation of Generated Multi-View Renderings}

\author[H. Liu, Y. Wang, V. G. Kim, \& M. Kazhdan]
{\parbox{\textwidth}{\centering
Hongyi Liu$^1$
\qquad
Yifan Wang$^2$
\qquad
Vladimir G. Kim$^3$
\qquad
Misha Kazhdan$^1$
}
\\
{\parbox{\textwidth}{\centering $^1$Johns Hopkins University, USA\quad$^2$Adobe Research, USA\quad$^3$Adobe Firefly Foundry, USA}
}
}

\begin{document}

\teaser{
 \includegraphics[width=0.9\textwidth]{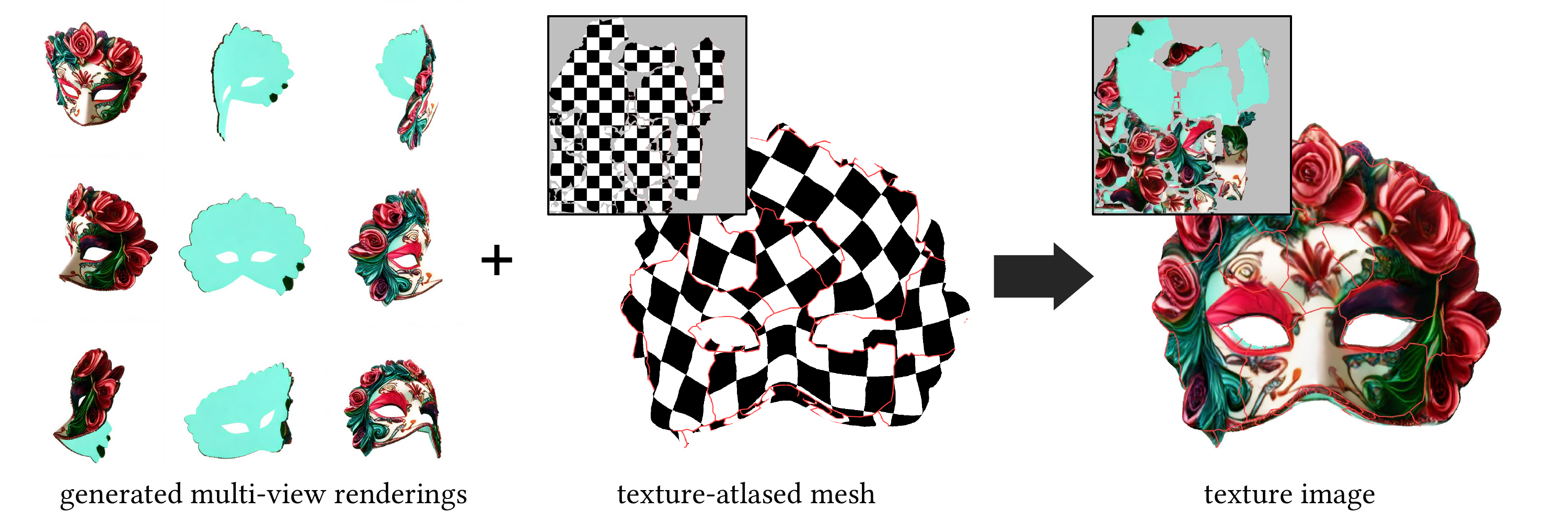}
 \centering
  \caption{Our system takes a small number of generated renderings with known camera parameters, together with a texture-atlased mesh, and generates a seamless texture image that resolves inconsistencies between the input renderings.}
\label{fig:teaser}
}

\maketitle
\begin{abstract}
We present a system for consolidating a sparse set of multi-view renderings into a single coherent texture image, providing a framework for integrating generative models for image generation within the geometry-processing pipeline. Given a texture-atlased triangle mesh and a set of renderings with associated camera parameters, we generate the texture image by: (1)~back-projecting the renderings onto the mesh to obtain texture images with normalized colors and confidences; (2)~solving for flow-fields on the mesh that align the colors to each other, taking into account the associated confidences; and (3)~fusing the registered colors into a single texture image using gradient-domain stitching, again incorporating confidences. Our approach complements existing techniques for multi-view image generation, leveraging mesh-based optical flow and texture-space stitching. In doing so it addresses the challenge that while modern image generation pipelines generate highly convincing renderings, they struggle to produce renderings that are consistent across views. Code is available at \url{https://github.com/hliuw/TextureConsolidation}.



\end{abstract}  

\section{Introduction}
The generation of 3D assets from generated multi-view renderings has advanced considerably in recent years~\cite{poole2023dreamfusion, TEXTure, Im2SurfTex}. Leveraging powerful diffusion-based models, these methods generate high-quality multi-view images for diverse 3D content. However, the renderings often exhibit inconsistencies when mapped onto 3D geometry, making it difficult to integrate them into a single, view-independent, model.

Motivated by this challenge, we consider the following problem: \textit{Given a set of multi-view renderings and a texture-atlased mesh, how do we construct a globally consistent texture image?}

This question has been explored for real-world RGB-D datasets, where approaches typically address the problem through view selection, pose optimization, registration, and blending. However, generated renderings differ from real-world captures in that they lack a coherent, view-independent model from which the renderings are generated, often resulting in geometric and photometric inconsistencies between renderings. As a result, while generating new views is computationally cheap, additional renderings have the down-side of introducing further inconsistency. This makes view augmentation a trade-off, in contrast to real-world captures where more data is preferred. In particular, this motivates the need for robust image registration and fusion -- supporting the generation of higher-quality textures from a \textit{small number} of renderings. %

While many multi-view texture registration methods exist, they are predominantly designed for dense RGB-D captures, relying on view selection to discard low-quality frames affected by motion blur, and having lighting and shading artifacts. generated multi-view renderings that sparsely sample the view sphere present a different set of challenges: each point is visible in only a few views; and generated views are not geometrically consistent, so that back-projection produces misaligned textures.

We address these challenges by designing a confidence-aware pipeline for multi-view texture consolidation that back-projects the renderings to texture images and then registration and stitching in the surface domain.
We extend existing optical flow \cite{prada2016motion} and gradient-domain stitching \cite{PoissonImageEditing,levin2003Image,prada2018} techniques, adapting them to support per-texel confidences.
This is achieved via quadrature-based finite-elements integration supporting sub-triangle evaluation of functions, thereby bridging texel-resolution and mesh-resolution discretizations.

Our method is agnostic to the rendering pipeline, serving as a general post-processing module that complements prior multi-view generation methods. Empirically, we show that existing registration methods for RGB-D captures degrade on generated inputs that sparsely sample the view sphere, while our framework produces high-quality texture images in this challenging setting.

\section{Related Work}
\paragraph*{Textured 3D Shape Generation} Diffusion models have achieved remarkable progress in 2D image generation \cite{rombach2022high, DiffusionSurvey, controlnet}. However, generating textured 3D models continues to be a challenge. Approaches differ in their choice of geometric representations and in how they couple geometry and color generation.

One common approach is to generate textured 3D shapes from user-provided text input. Approaches using Neural Radiance Fields (NeRF) \cite{Mildenhall:2020:ECCV} or Gaussian-Splats (GS) \cite{Kerbl:TOG:2023} often start by either directly optimizing the scene from text prompts, or leveraging diffusion models to first generate multi-view renderings which are then used to reconstruct the NeRF/GS model \cite{poole2023dreamfusion, wang2023prolificdreamer, Lin_2023_CVPR,tang2024lgm, tang2023dreamgaussian, yi2023gaussiandreamer}.
Hybrid scene representations have also been used to transition between different intermediate representations during optimization before extracting the textured mesh \cite{chen2023fantasia3d}. Additionally, some NeRF/GS methods enable mesh extraction as a final step \cite{Lin_2023_CVPR, tang2023dreamgaussian, wei2025meshlrm}.

Another approach takes an input 3D geometry and generates color textures by back-projecting multi-view renderings \cite{TEXTure, chen2023text2tex, paint3d, Im2SurfTex, cheng2024mvpaint, enforce}. These methods first generate multi-view rendering from pretrained 2D diffusion models, often using depth maps as guidance. The composite color texture is then produced through aggregation of multi-view renderings and progressive inpainting of untextured regions. Our work focuses on this setting and addresses the challenge of registering and compositing such multi-view renderings.

\paragraph*{Texture Registration and Blending} A challenge in back-projecting multi-view renderings is the lack of consistency across different views -- a long-standing challenge in 3D reconstruction from photographs. Such inconsistencies typically arise from variations in lighting, occlusions, and geometric distortions. generated textures can exacerbate the problem by independently introducing different details for different renderings. %

Addressing these issues typically involves some form of registration and blending. A common strategy is to align textures through warping. \citeauth{Eisemann}{Eisemann08FT} and \citeauth{Casas}{4Dtexture} use reprojected images to compute a correction field via optical flow. \revise{\citeauth{Dellepiane}{dellepiane2012} compute optical flow between overlapping photographs to locally correct image-to-geometry misalignment.} \citeauth{Aganj}{impreciseMesh} compute the displacement using reprojected feature correspondences across interest points in different views. Inconsistency can also be mitigated by optimizing camera pose. Zhou and Koltun \cite{CMO} jointly optimize camera poses and non-rigid deformation per photo to maximize photometric consistency across projected color images. \citeauth{Huang}{huang2020adversarial} introduce adversarial texture optimization that learns tolerance to misalignments by training a discriminator on view projections, later extended by \citeauth{Zhao}{adv_align} using Markov Random Field view initialization and FFT alignment. \citeauth{Fu}{fu2021seamless} incorporate a texture detail layer to select the best-fit texture for each triangle face. Camera pose is then optimized to align these renderings, and they are regenerated at the texture charts' boundaries to remove seams. \citeauth{Wang}{SeamlessAligned} propose a similar method by detecting and excluding misaligned regions, optimizing camera positions, and filling in holes. \revise{Kim et al.~\cite{kim2026robust} refine an initial texture by jointly optimizing uncertainty maps and deformation grids to handle inconsistent AI-generated targets within non-differentiable production engines.}

Other methods focus on view blending. This is often done as a combination of view selection and stitching. \revise{An early example is \citeauth{Gal}{gal2010},
which assigns each face both a source image and a local image-space
translation compensating for misalignment.} \citeauth{Waechter}{mvs-tex} texture meshes from photos by filtering occluders via photo-consistency checks, selecting views based on gradient magnitude, and performing color adjustment using gradient-domain stitching \cite{PoissonImageEditing}. \revise{\citeauth{Pagés}{pages2015} blend the contributions of all cameras across the whole mesh, using smoothly interpolated per-vertex camera ratings to suppress seams while limiting blur.} \citeauth{Du}{montage4D} instead focus on texture blending, identifying seams using rendered depth maps and modulating them using geodesic fields. These methods are also applicable to generated multi-view renderings. \citeauth{Zhao}{enforce} select a subset of generated renderings to minimize inconsistency via combinatorial optimization and warp them using SIFT flow \cite{SIFTflow}. The resulting textures are stitched by assigning each face an optimal projected texture, formulated as a joint labeling problem. Im2SurfTex \cite{Im2SurfTex} introduces a neural back-projection module that integrates textures by encoding geodesic-aware features and fusing neighboring pixels through cross-attention.

In our approach we use optical flow to register the textures and gradient-domain stitching to composite them, performing both over the surface. This complements earlier methods that perform view-dependent processing and rely on a final blending step to consolidate results. We provide a brief review of these techniques.

\paragraph*{Optical Flow} Optical flow techniques were originally developed to identify the motion of objects in consecutive image frames by estimating a vector-field under the assumption of luminance constancy \cite{HORN1981, opticalflowsurvey, alfarano2024estimating}. They have also been used to register textures from multi-view renderings by computing an aligning flow-field and advecting the pixels \cite{Eisemann08FT, dellepiane2012, 4Dtexture}. Traditionally, though, these methods work in the (flat) 2D image domain.%

Extending optical flow to the (curved) surface domain requires evaluating the smoothness of tangent vector-fields \cite{de2016vector, lee2018introduction}. This is achieved by defining a Dirichlet energy for tangent vector-fields \cite{sharp2019vector, stein2020simple, brandt2017spectral}. 
With this energy, the optical flow-field can be computed using a variational formulation \cite{lefevre2008optical, prada2016motion}. Signals are then advected along the flow-field to align details and reduce the ghosting artifacts that occur with naive blending.%

\begin{figure*}[h]
  \centering
    \includegraphics[width=.95\textwidth]{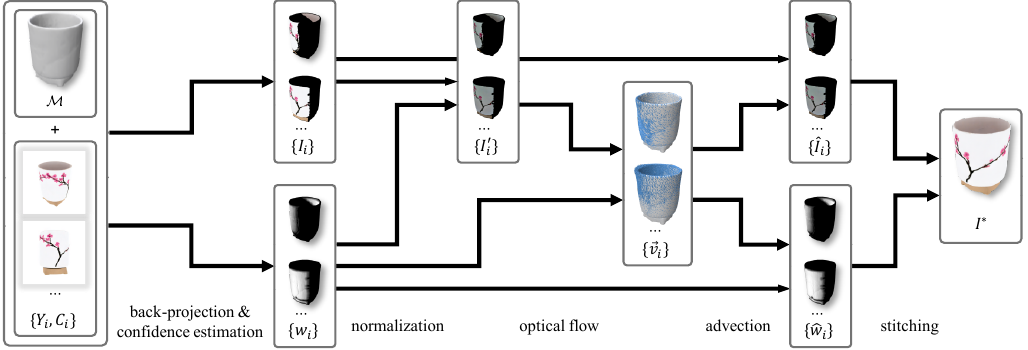}
    \caption{Starting with a texture-atlased mesh and multi-view renderings with associated camera parameters ($\Surface$ and $\{\Render_i,\Camera_i\}$), we back-project the renderings onto the mesh to get initial texture color and confidence images ($\{\Color_i\}$ and $\{\Confidence_i\}$). We compute the normalized colors ($\{\NormalizedColor_i\}$), adjusting for lighting variations. Taking into account confidences, we register the normalized colors to each other, obtaining a flow-field for each color texture ($\{\FlowField_i\}$). We advect the initial back-projected colors and confidences using the flow-field ($\{\AdvectedColor_i,\AdvectedConfidence_i\}$). Finally, we perform gradient-domain stitching to get a seamless texture image ($\OutputColor$).
    }
    \label{fig:approach}
\end{figure*}

\paragraph*{Gradient-domain Stitching} A popular method for seamless stitching fuses together signals by compositing their gradients, rather than colors. The approach was initially formulated for 2D images \cite{levin2003Image, PoissonImageEditing} but has since been applied to the stitching of signals over a triangulated surface \cite{arnaud2014,Chuang2016}. More recently, the approach has been extended to operate within the texture domain, handling cross-chart discontinuities and non-uniform distortion in the texture-mapping \cite{prada2018}.

\section{Overview}
\label{sec:overview}
\begin{comment}
\begin{table}[h]
\centering
\begin{tabular}{@{}cl@{}}
\toprule
Symbol        & Meaning  \\ \midrule
$\mathcal{M}$ & triangle mesh \\
$\{\Render_i\}$       & renderings \\
$\{\Camera_i\}$      & camera parameters \\
$\{\Color_i\}$       & back-projected colors \\
$\{\Confidence_i\}$       & confidences \\
$\{\NormalizedColor_i\}$ & normalized colors \\
$\{\AdvectedColor_i\}$ & advected colors \\
$\{\AdvectedConfidence_i\}$ & advected confidences \\
{\FlowFieldi}\{\FlowField_i\} & optical flow-fields \\
\NormalizationWeight\NormalizationWeight & normalization parameter \\
\ScalarSmoothingWeight\ScalarSmoothingWeight & initial scalar smoothing weight \\
\FlowFieldSmoothingWeight\FlowFieldSmoothingWeight & flow-field smoothing weight \\
\GradientFittingWeight\GradientFittingWeight & gradient fitting weight \\
\GradientDomainRegularizationWeight\GradientDomainRegularizationWeight & gradient-domain regularization weight \\
\OutputColor\OutputColor & output texture \\
\bottomrule
\end{tabular}
\caption{Notation}
\label{tab:notation}
\end{table}
\end{comment}

The input to our method is a texture-atlased manifold triangle mesh $\Surface\subset\mathbb{R}^3$ and a sparse set of multi-view renderings with corresponding camera parameters, $\big\{ \{\Render_i,\Camera_i\}\big\}_{i=1}^K$. Our goal is to produce a globally consistent texture image, $\OutputColor$. This is achieved through the pipeline described below and visualized in \figureref{fig:approach}.

\textit{Preprocessing.}
We generate texture color and confidence images for each input rendering -- back-projecting each rendering onto the surface, defining per-texel confidence from visibility, depth, and viewing angle, and normalizing the colors.

\textit{Registration.}
We align colors in the surface domain -- computing an optical flow-field for each color and confidence image, and advecting texel values along the fields.

\textit{Stitching.}
We fuse the optical-flow aligned texture color and confidence images -- performing confidence-aware stitching in the gradient-domain.

\begin{figure}[h]
        \centering
        \begin{subfigure}[b]{0.15\textwidth}
            \centering
            \includegraphics[width=\textwidth]{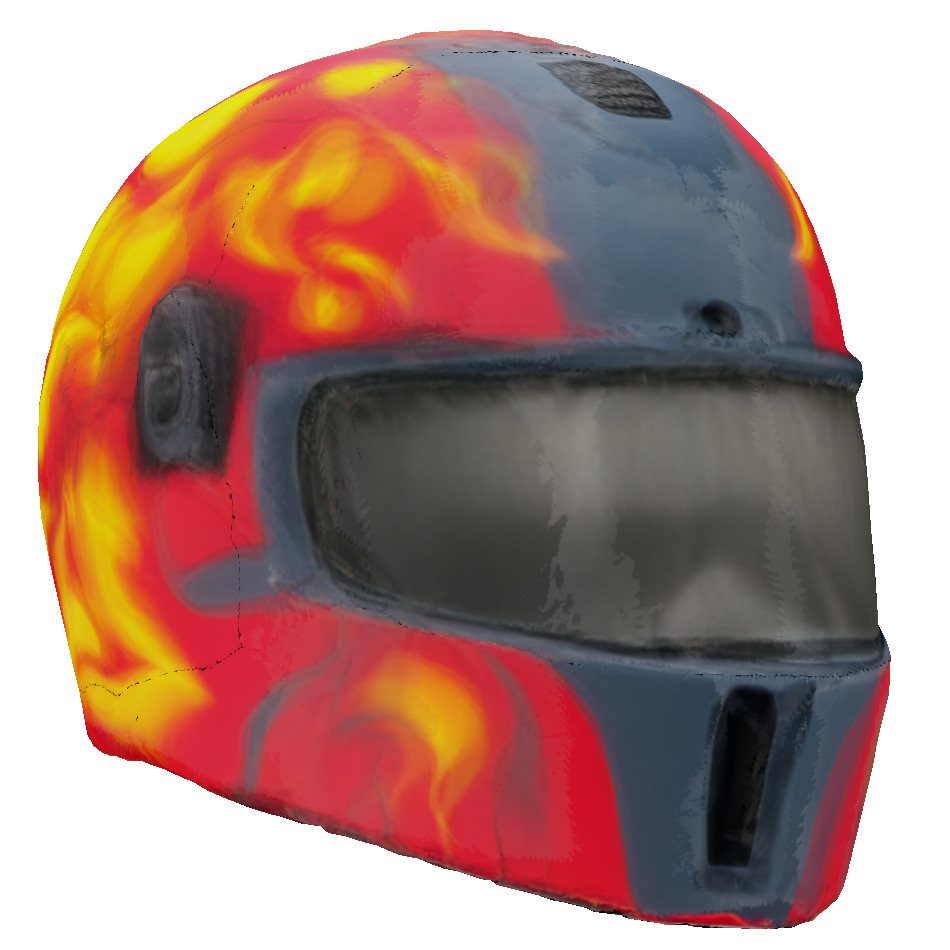}
        \end{subfigure}
        \begin{subfigure}[b]{0.15\textwidth}  
            \centering 
            \includegraphics[width=\textwidth]{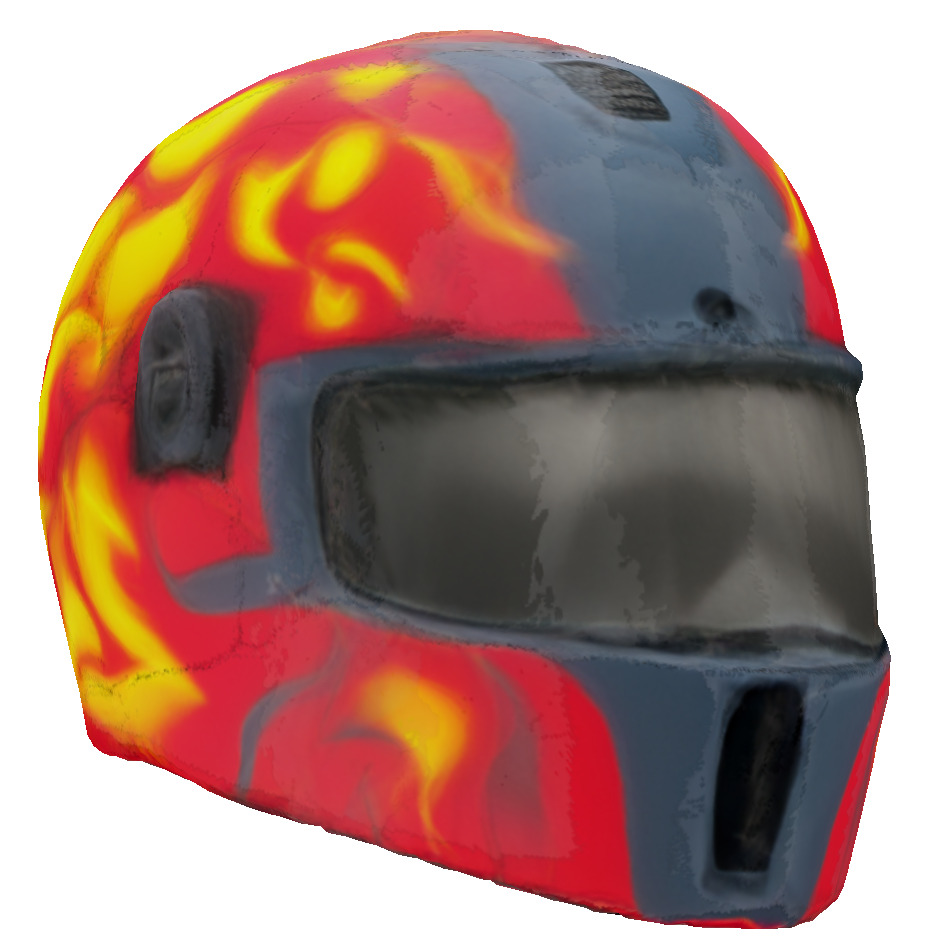}
        \end{subfigure}
        \begin{subfigure}[b]{0.15\textwidth}   
            \centering 
            \includegraphics[width=\textwidth]{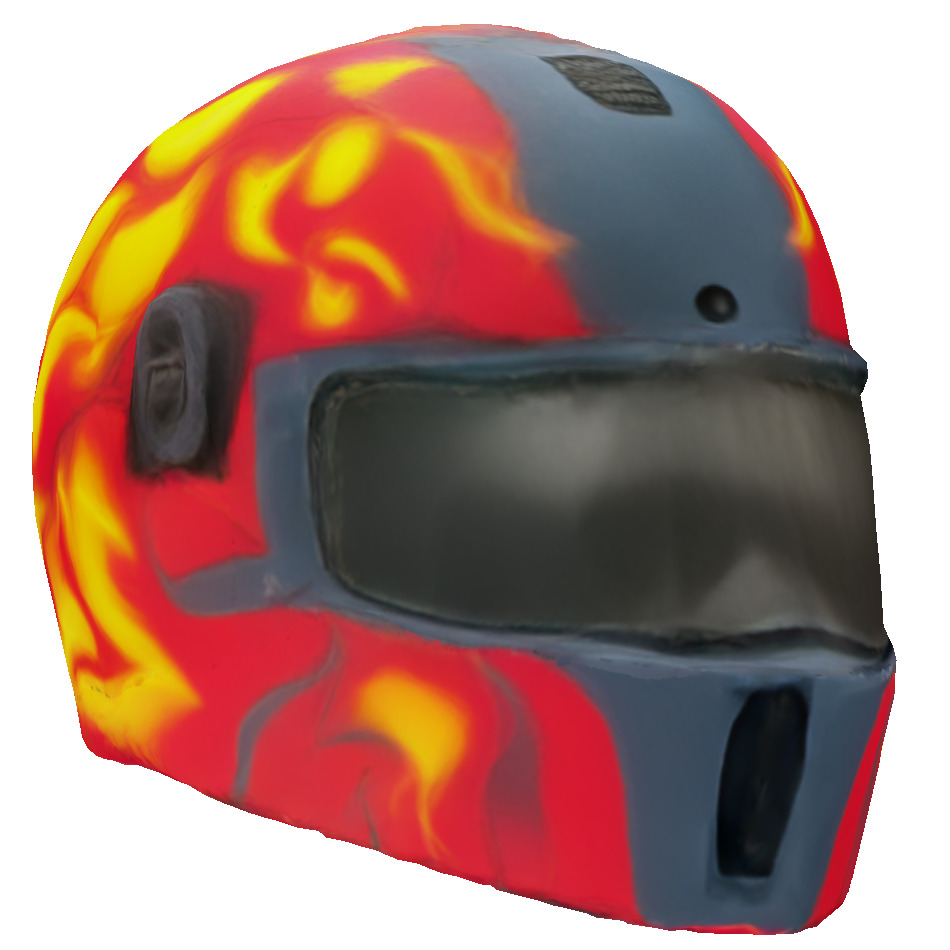}
        \end{subfigure}
        \begin{subfigure}[b]{0.15\textwidth}
            \centering
            \includegraphics[width=\textwidth]{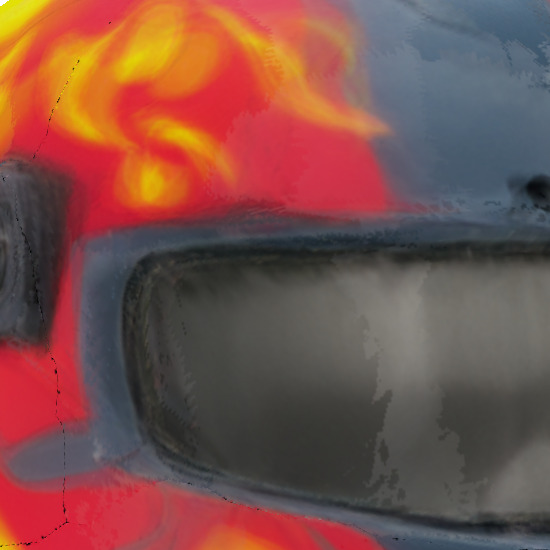}
             \caption*{blended}     
        \end{subfigure}
        \begin{subfigure}[b]{0.15\textwidth}  
            \centering 
            \includegraphics[width=\textwidth]{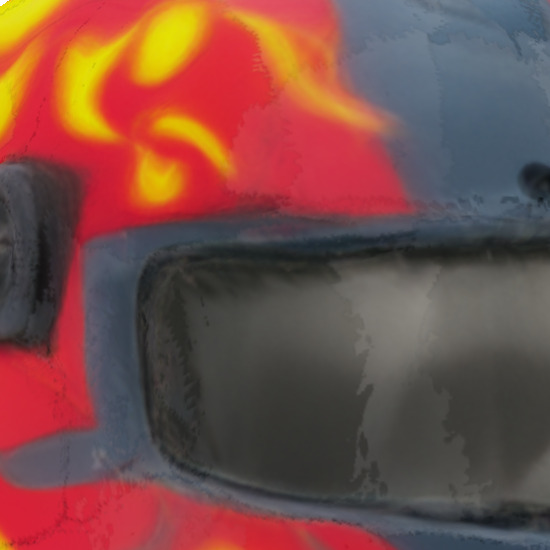}
             \caption*{registered \& blended}    
        \end{subfigure}
        \begin{subfigure}[b]{0.15\textwidth}   
            \centering 
            \includegraphics[width=\textwidth]{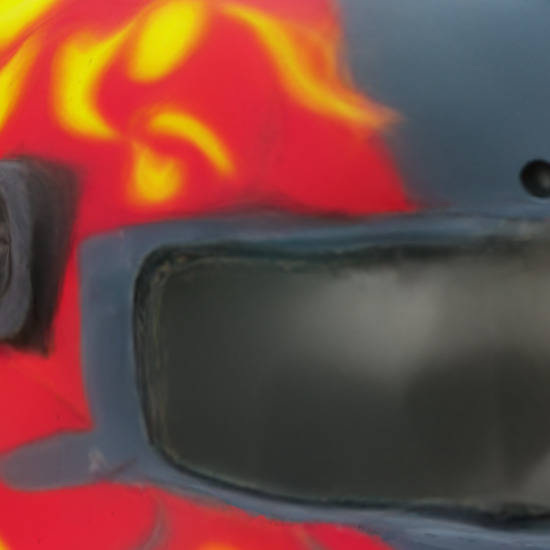}
             \caption*{registered \& stitched}
        \end{subfigure}
        \caption{
        Confidence-weighted blending produces ghosting due to feature misalignment (left). Optical flow aligns the features, but discontinuities due to color variations persist (middle). Gradient-domain stitching removes these, producing a seamless texture image without ghosting (right).
        }
        \label{fig:pipeline_comparison}
\end{figure}

\figureref{fig:pipeline_comparison} motivates the need for both registration and stitching. Just blending the back-projections using confidences creates a texture with ghosting, due to the misalignment of color features (left). Optical flow addresses the ghosting issue by aligning the back-projections to each other, but still exhibits discontinuities due to color variations across the input renderings (center). Stitching produces a seamless and ghost-free color texture (right).
\section{Preliminaries}
\label{sec:preliminaries}

We review gradient-domain processing and optical flow computation, for signals defined over a mesh $\Surface\subset\R^3$. In what follows, we refer to functions and vector-fields as \textit{queryable} if they are bounded and can be evaluated almost anywhere on the mesh. That is, if they can be integrated over the mesh.

\subsection{Gradient-Domain Processing}
The goal of classical gradient-domain processing is to solve for a signal $f$ that simultaneously fits target values $h$ and gradients $\FlowField$:
\begin{equation}
E_{\text{gd}}(f,h,\vec{v},\GradientFittingWeight)=\SurfaceIntegral(f-h)^2+\GradientFittingWeight\|\nabla f-\FlowField\|^2
\label{eq:gradient-domain}
\end{equation}
with $\GradientFittingWeight$ the gradient-fitting weight.

Two standard applications are smoothing and stitching. In smoothing, $\vec{v}$ is set to zero and $\GradientFittingWeight$ parametrizes the ``extent'' of the smoothing. In stitching, $\vec{v}$ is set to the composite gradients of the input images to be stitched, and $\GradientFittingWeight$ is set to a large value -- focusing the optimization on fitting the gradients.

\paragraph*{Discretization} Given a basis of scalar functions $\{\ScalarBasis_i\}_{i=1}^N$ on $\Surface$,
the problem reduces to the solution of a sparse symmetric positive definite system, with solution coefficients, $\mathbf{x}\in\R^N$, satisfying:
$$(\mathbf{M}+\GradientFittingWeight\cdot\mathbf{S})\cdot\mathbf{x}=\mathbf{b}$$
with $\mathbf{M},\mathbf{S}\in\R^{N\times N}$ the mass and stiffness matrices, and $\mathbf{b}\in\R^N$ encoding the scalar and gradient constraints:
\begin{gather*}
\mathbf{M}_{ij}\equiv\SurfaceIntegral\ScalarBasis_i\cdot\ScalarBasis_j,\quad \mathbf{S}_{ij}\equiv\SurfaceIntegral\langle \nabla\ScalarBasis_i,\nabla\ScalarBasis_j\rangle,\,\,\,
\\
\mathbf{b}_i
\equiv\SurfaceIntegral\ScalarBasis_i\cdot h + \GradientFittingWeight\langle\nabla\ScalarBasis_i,\vec{v}\rangle.
\end{gather*}
Essential to our approach is that, using quadrature to approximate integrals, defining the linear system only requires that the $\ScalarBasis_i$ and their derivatives, as well as the constraints $h$ and $\vec{v}$, be queryable. 

\subsection{Optical Flow}
Given source and target signals, $f_s,f_t:\Surface\rightarrow\R^d$, the goal of optical flow is to compute a tangent vector-field describing how the source should advect along the mesh so as to align with the target.

The approach uses the fact that when a signal $\psi$ advects along a vector-field $\FlowField$, the advected value at a point $p\in\Surface$ is ``brought in'' from the direction opposite the flow -- i.e. the direction from which the flow arrives. Formally, the time-evolving signal obtained by advecting along $\FlowField$ satisfies the PDE \cite{Mullen:FCM:2011}:
$$\frac{\partial\psi}{\partial t} =-\langle\FlowField,\nabla\psi\rangle.$$
Thus, for small time-steps $\epsilon$, the advected signal is approximately:
$$\hbox{Adv}_{\epsilon\cdot\FlowField}(\psi)\approx\psi - \epsilon\cdot\langle\FlowField,\nabla\psi\rangle.$$
This gives an optical flow energy measuring the difference between the (approximately) advected source and the target:
\begin{align}
E_{\text{of}}(\FlowField,f_s,f_t,\FlowFieldSmoothingWeight)
&= \SurfaceIntegral\Big(\big(f_s-\langle\FlowField,\nabla f_s\rangle\big)-f_t\Big)^2 + \FlowFieldSmoothingWeight\|\nabla\FlowField\|^2.%
\label{eq:optical-flow}
\end{align}
The term $\FlowFieldSmoothingWeight\|\nabla\FlowField\|^2$ acts as a regularizer to address the under-determinedness, encouraging smoother flow-fields \cite{HORN1981}.

As the solution $\FlowField$ should itself be time-varying, the optimization is iterated -- at each time-step the previously estimated flow-fields are used to advect the source and a correcting flow-field is computed to register the advected source to the target.

\paragraph*{Implementation}
We follow the implementation of \citeauth{Prada}{prada2016motion} which performs the time-stepping hierarchically -- initially solving for a smooth flow-field that aligns the low-frequency components and progressively solving for higher-frequencies. This is achieved by modifying the input signals and the regularization in \equationref{eq:optical-flow} as a function of the hierarchy level. The advected source and target are replaced by their blurred versions, with the amount of blurring reduced at later levels of the hierarchy. Similarly, the smoothness regularization weight $\FlowFieldSmoothingWeight$ is decreased at later levels.

In principle, the deformation should be expressed as the composition of advections along the iteratively estimated flow-fields. However, both within the algorithm and as its output, Prada~\etal{} propose using a single flow-field that is their sum.

\paragraph*{Discretization}
Given a vector-field basis, $\{\VectorBasis_i\}_{i=1}^N$,
the problem reduces to the solution of a sparse symmetric positive definite system, with the coefficients of the flow-field, $\mathbf{x}\in\R^N$, satisfying:
$$(\mathbf{M}^{\nabla f_s}+\FlowFieldSmoothingWeight\cdot\mathbf{S})\cdot\mathbf{x}=\mathbf{b}$$
where $\mathbf{M}^{\nabla f_s},\mathbf{S}\in\R^{N\times N}$ are the $\nabla f_s$-weighted mass and stiffness matrices, and $\mathbf{b}\in\R^N$ encodes the difference constraints:
\begin{gather*}
\mathbf{M}^{\nabla f_s}_{ij}\equiv\SurfaceIntegral\langle\VectorBasis_i,\nabla f_s\rangle\cdot\langle\VectorBasis_j,\nabla f_s\rangle,\quad
\mathbf{S}_{ij}\equiv\SurfaceIntegral\langle \nabla\VectorBasis_i,\nabla\VectorBasis_j\rangle,\\
\mathbf{b}_i\equiv\SurfaceIntegral\langle\VectorBasis_i,\nabla f_s\rangle\cdot(f_t-f_s).
\end{gather*}
As above, using quadrature only requires that the $\VectorBasis_i$, their derivatives, the signals $f_s$ and $f_t$, and the gradient $\nabla f_s$ be queryable.

\paragraph*{Signal Advection}
Finally, given a signal $f$ and a flow-field $\FlowField$, the advected signal, $\hbox{Adv}_{\FlowField}(f)$, is obtained by stepping backwards along the vector-field, using the hinge-map \cite{de2016vector} to flatten adjacent triangles when stepping across an edge, and sampling.

\section{Quadrature-based Integration}
\label{sec:quadrature}
Discretizing the computation of optical flow requires resolving two conflicting challenges. On the one hand, the flow-field is expected to be relatively low-frequency, corresponding to the fact that the underlying deformation is smooth. On the other, we would like to represent the colors and confidences at the finer resolution of texel space. Additionally, both for computing the flow-field and stitching, we need the implementations to adapt to confidence.

We resolve these challenges using quadrature-based integration, which also supports mixing different signal representations. For scalar functions, we use the mesh-resolution ``hat'' basis and the texel-resolution basis of \citeauth{Prada}{prada2018}. For vector-fields we use the mesh-resolution basis functions of \citeauth{Liu}{liu2026}. %

Using quadrature makes it straight-forward to incorporate color and confidence represented at  texel-resolution when discretizing the flow-field at the vertex resolution. For completeness, we step through the implementation. In particular, we describe the evaluation of integrals for a triangle $\tau=\{p_0,p_1,p_2\}\subset\R^3$, with the actual computation performed over the unit right triangle:
$$\T=\{(s,t)\in[0,1]^2|s+t\leq1\}.$$
We assume that we are given a confidence map $w:\T\rightarrow[0,1]$, we set the metric tensor to be $\textbf{g}\in\R^{2\times 2}$ with $\textbf{g}_{ij}=\langle p_i-p_0,p_j-p_0\rangle$, and we denote its determinant $|\textbf{g}|$.
Lightly abusing notation, for a scalar- (resp. vector-)valued function $f:\T\rightarrow\R$ (resp. $\vec{v}:\T\rightarrow\R^2$), we write ``$f$'' (resp. ``$\vec{v}$'') to represent both the function on the unit right triangle $\T$ and the function on the embedded triangle $\tau\subset\R^3$.%

\subsection{Integrating Scalar Functions}
Given a scalar basis function $\ScalarBasis_i:\T\rightarrow\R$ and some other function $f:\T\rightarrow\R$, we approximate the integral of the confidence-weighted dot-product of $\ScalarBasis_i$ and $f$ as:
\begin{equation}
\int_\tau\Confidence\cdot\ScalarBasis_i\cdot f
\approx\sqrt{|\textbf{g}|}\cdot\sum_{j=1}^{Q}\mu_j\cdot\Confidence(q_j)\cdot\ScalarBasis_i(q_j)\cdot f(q_j)
\label{eq:scalar-confidence-integral}
\end{equation}
with $\{q_j,\mu_j\}\subset\T\times\R^{\geq0}$ the quadrature locations and weights \cite{Cowper:1973:IJNME,Taylor:JCAM:2008}. This estimate of the integral only requires that the target and confidence functions, $f$ and $\Confidence$, be queryable, which is true for texture-mapped signals. A similar approach can be used to define the confidence-weighted mass and (assuming queryable derivatives of $\ScalarBasis_i$) stiffness matrices.

\subsection{Integrating Gradients of Scalar Functions}
Given a tangent vector-field basis function $\VectorBasis_i:\T\rightarrow\R^2$ and some other function $f:\T\rightarrow\R$, we approximate the confidence-weighted integral of the dot-product of $\VectorBasis_i$ and the gradient of $f$ as:
\begin{equation}
\int_\tau\Confidence\cdot\langle\VectorBasis_i,\nabla f\rangle\approx\sqrt{|\textbf{g}|}\cdot\sum_{j=1}^{Q}\mu_j\cdot\Confidence(q_j)\cdot\langle\VectorBasis_i(q_j),\nabla_{\mathbf{g}} f(q_j)\rangle_{\mathbf{g}}.
\label{eq:vector-confidence-integral}
\end{equation}
We note that the inner-product is equivalent to the pairing of the tangent vector-field $\VectorBasis_i$ with the differential of $f$, which is metric-independent.
As above, the incorporation of confidence only requires that the functions $f$, $\nabla f$, and $\Confidence$ be queryable. This too holds for texture-mapped signals. And again, a similar approach can be used to define the confidence-weighted mass and (assuming queryable derivatives of $\VectorBasis_i$) stiffness matrices.

\section{Method}
We now discuss the key steps of our pipeline, summarized in \algoref{alg:process}. To facilitate the presentation, we begin by describing how we adapt the traditional energies to incorporate confidence.

    \paragraph*{Gradient-Domain}
Given a set of target values, target gradients, and confidences, $\{h_i,\vec{v}_i,\Confidence_i\}$, we define the gradient-domain energy:
\begin{align}
\nonumber
&E_{\text{gd}}(f,\{h_i,\vec{v}_i,\Confidence_i\},\GradientFittingWeight,\GradientDomainRegularizationWeight)\\
&\quad\quad\quad\equiv\sum_i\SurfaceIntegral \Confidence_i\cdot\Big((f-h_i)^2+\GradientFittingWeight\|\nabla f-\vec{v}_i\|^2\Big)+\GradientDomainRegularizationWeight\|\nabla f\|^2
\label{eq:confidence-weighted-gradient_domain}
\end{align}
The additional regularization term, $\GradientDomainRegularizationWeight\|\nabla f\|^2$, is needed to make the energy non-singular in regions of zero confidence. (In general, $\GradientDomainRegularizationWeight$ is set to a very small value and serves solely to encourage the signal to be smooth in regions with no confidence.)

In contexts where we have multiple weight functions, we normalize and rescale the weights:
$$
w_i(q)\leftarrow 
\underbrace{\left(\frac{w_i(q)}{\sum_j w_j(q)}\right)\vphantom{\left(1-\prod_j(1-w_j(q))\right)}}_{\mbox{\scriptsize normalization}}\cdot\underbrace{\left(1-\prod_j(1-w_j(q))\right)}_{\mbox{\scriptsize scaling}}.
$$
Thus, the sum of the weights is $\left(1-\prod_j(1-w_j)\right)\in[0,1]$, giving the confidence that the point is covered by some target. (When all weights are zero, the normalized and rescaled weights remain zero.) 

\textit{Blurring.}
The energy associated with blurring a signal $h$ is obtained by setting $\GradientFittingWeight=0$ (and is independent of target gradient):
\begin{equation}
E_\text{blur}(f,h,w,\GradientDomainRegularizationWeight)\equiv\SurfaceIntegral w\cdot(f-h)^2+\GradientDomainRegularizationWeight\|\nabla f\|^2
\label{eq:confidence-weighted-heat-diffusion}
\end{equation}

\textit{Stitching.}
In the context of stitching together individual images, most often the target signals are the images themselves and the target vector-fields are their gradients, $\vec{v}_i=\nabla h_i$. This gives:
\begin{equation}
E_{\text{stitch}}(f,\{h_i,\Confidence_i\},\GradientFittingWeight,\GradientDomainRegularizationWeight)\equiv E_\text{gd}(f,\{h_i,\nabla h_i,\Confidence_i\},\GradientFittingWeight,\GradientDomainRegularizationWeight).
\label{eq:confidence-weighted-stitching}
\end{equation}
This confidence-weighted expression obviates the challenges of compositing colors/gradients into a single function/vector-field, typically required as input for stitching. The confidences $\Confidence_i$ implicitly dictate which values/gradients (and to what extent) should be ``active'' in different parts of the domain.

    \paragraph*{Optical Flow}
Given source and target signals $f_s$ and $f_t$, and confidence $\Confidence$, the optical flow energy is adapted as:
\begin{equation}
E_{\text{of}}(\FlowField,f_s,f_t,w,\FlowFieldSmoothingWeight)
\equiv\SurfaceIntegral w\cdot\Big((f_s-\langle\FlowField,\nabla f_s\rangle) - f_t)\Big)^2 + \FlowFieldSmoothingWeight\|\nabla\FlowField\|^2.
\label{eq:of_energy}
\end{equation}

    \paragraph*{Optimization}
For the above energies, we denote their minimizers:
\begin{align}
\nonumber
\text{Stitch}(\{h_i,\Confidence_i\},\GradientFittingWeight,\GradientDomainRegularizationWeight)&\equiv\argmin_{f} E_\text{stitch}(f,\{h_i,\Confidence_i\},\GradientFittingWeight,\GradientDomainRegularizationWeight)\\
\nonumber
\text{Blur}(h,w,\GradientDomainRegularizationWeight)&\equiv\argmin_{f} E_\text{blur}(f,h,w,\GradientDomainRegularizationWeight)\\
\text{Flow}(f_s,f_t,w,\FlowFieldSmoothingWeight)&\equiv\argmin_{\FlowField} E_\text{of}(\FlowField,f_s,f_t,w,\FlowFieldSmoothingWeight).
\label{eq:confidence-weighted-solutions}
\end{align}
Given a choice of queryable scalar/vector-field basis functions, the construction of the associated linear systems directly parallels the process described in \sectionref{sec:preliminaries}, with integration adapted to incorporate confidence, as in \equationsref{eq:scalar-confidence-integral} and~\ref{eq:vector-confidence-integral}. 

\subsection{Preprocessing}
Our pre-processing consists of two stages. Back-projecting to turn the renderings into texture color and confidence images, and normalization to correct for low-frequency variation of color in the generated rendering.

\subsubsection*{Back-Projection}

\begin{figure}
    \centering
    \includegraphics[width=\linewidth]{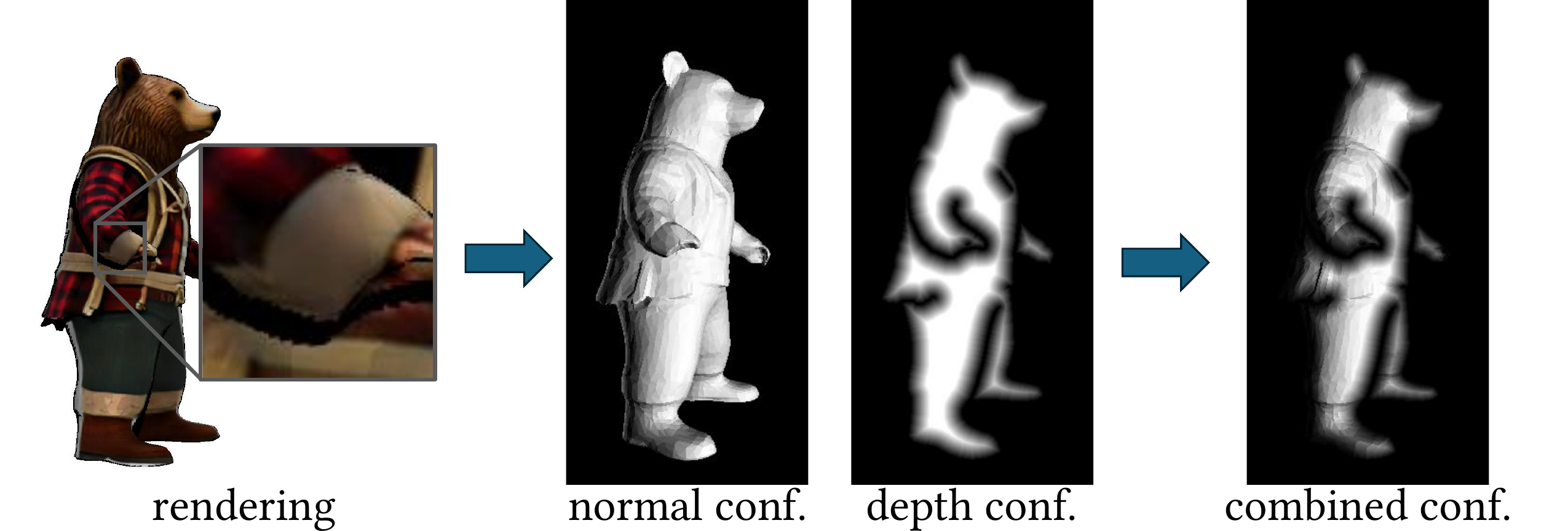}
    \caption{At silhouettes, geometric inconsistency causes back-projection to map colors to the wrong location (left). Defining confidences that account for alignment with the camera's view-direction and depth discontinuities (center), we give less weight to texels projecting near silhouettes (right).}
    \label{fig:confidence}
\end{figure}

Given the renderings and camera parameters, $\big\{\{\Render_i,\Camera_i\}\}_{i=1}^K$, we back-project the renderings  onto the mesh $\Surface$ to obtain color and confidence textures, $\big\{\{\Color_i,\Confidence_i\}\big\}_{i=1}^K$.

    \paragraph*{Color}
Given a rendering/camera $\{\Render_i,\Camera_i\}$, we define the color at a texel by locating the corresponding point on the mesh, back-projecting using the camera parameters, and sampling $\Render_i$ at the back-projected position. The surface point is computed in two steps. First, we compute the barycentric coordinates of the texel with respect to the closest texture triangle. Then, we compute the barycentrically weighted average of the triangle's corner positions in 3D. When the texel falls inside a texture triangle, the barycentric weights are positive and the average of the corners lies on the mesh. Otherwise, we map the average to the closest point on the mesh.

    \paragraph*{Confidence}
Because the generated renderings may have silhouettes that are not consistent with the input geometry, back-projection can cause colors to fall far from their desired location on the surface. We address this by defining a confidence texture that combines normal and depth information. (\figureref{fig:confidence} shows an example for the Bear model, where part of the belt near the silhouette of the arm is back-projected to the wrist. Combining normal and depth information, we assign texels near silhouettes lower confidence.)

\textit{Normal Confidence.}
We assign normal confidences by measuring the extent to which texels are front-facing to the camera. These are computed in the standard way, using the clamped dot-product of the surface normal and the camera's viewing direction. Though this identifies locations on the silhouette, it fails to identify front-facing background pixels that are back-projected to the foreground.

\textit{Depth Confidence.}
To assign lower confidence to texels back-projecting to the foreground \textit{near} silhouettes, we first identify pixels where the depth-map is discontinuous. This is achieved by rendering the scene from the perspective of the camera, computing the depth-map, and thresholding its Laplacian. The confidence is set to zero where the Laplacian exceeds the threshold and is reduced nearby. The confidence of occluded texels is also set to zero.

Finally, we multiply the depth and normal confidences and quantize low-confidences to zero. (See \appendixref{supp:parameters} for more details.)

While we instantiate these weights using normal/view-direction alignment and depth continuity, our approach equivalently accommodates other confidence information such as rendering quality/blurriness. %
Since  our primary focus is registration and stitching, this step serves mainly to produce meaningful texture color and confidence images suitable for down-stream processing.

\subsubsection*{Color Normalization}
We adjust the input signals to eliminate low-frequency color variations before estimating the flow. This is done by normalizing with respect to local estimates of the mean and scale:
\begin{equation}
\label{eq:normalization}
\NormalizedColor_i\leftarrow\frac
{\Color_i - \text{Blur}(\Color_i,\Confidence_i,\NormalizationWeight)}
{\text{Blur}\Big(\left\|\Color_i-\text{Blur}(\Color_i,\Confidence_i,\NormalizationWeight)\right\|,\Confidence_i,\NormalizationWeight\Big)}.
\end{equation}
The blurring is done taking into account confidence and $\NormalizationWeight>0$ is a parameter describing the extent of the region over which the average and scale are estimated. %

\begin{figure}
    \centering
    
    \begin{subfigure}[b]{\columnwidth}
        \centering
        \includegraphics[width=0.35\textwidth]{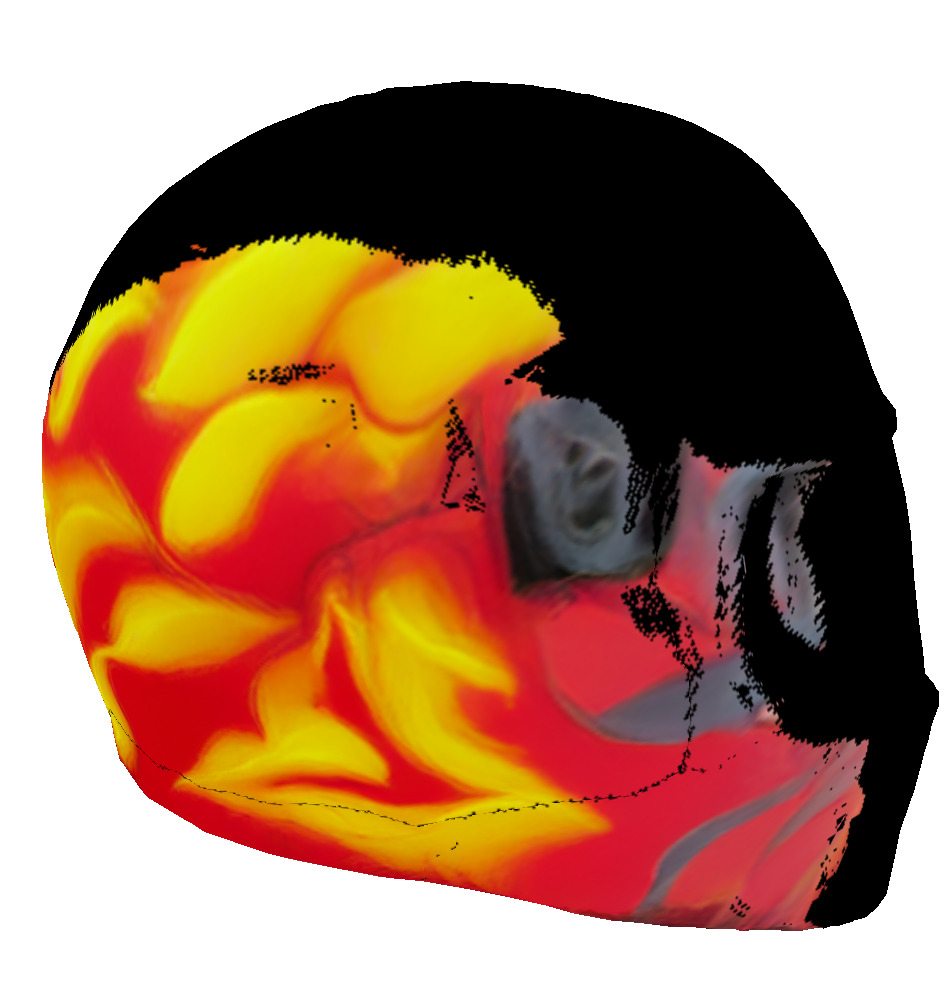}
        \hspace{0.05\textwidth}
        \includegraphics[width=0.35\textwidth]{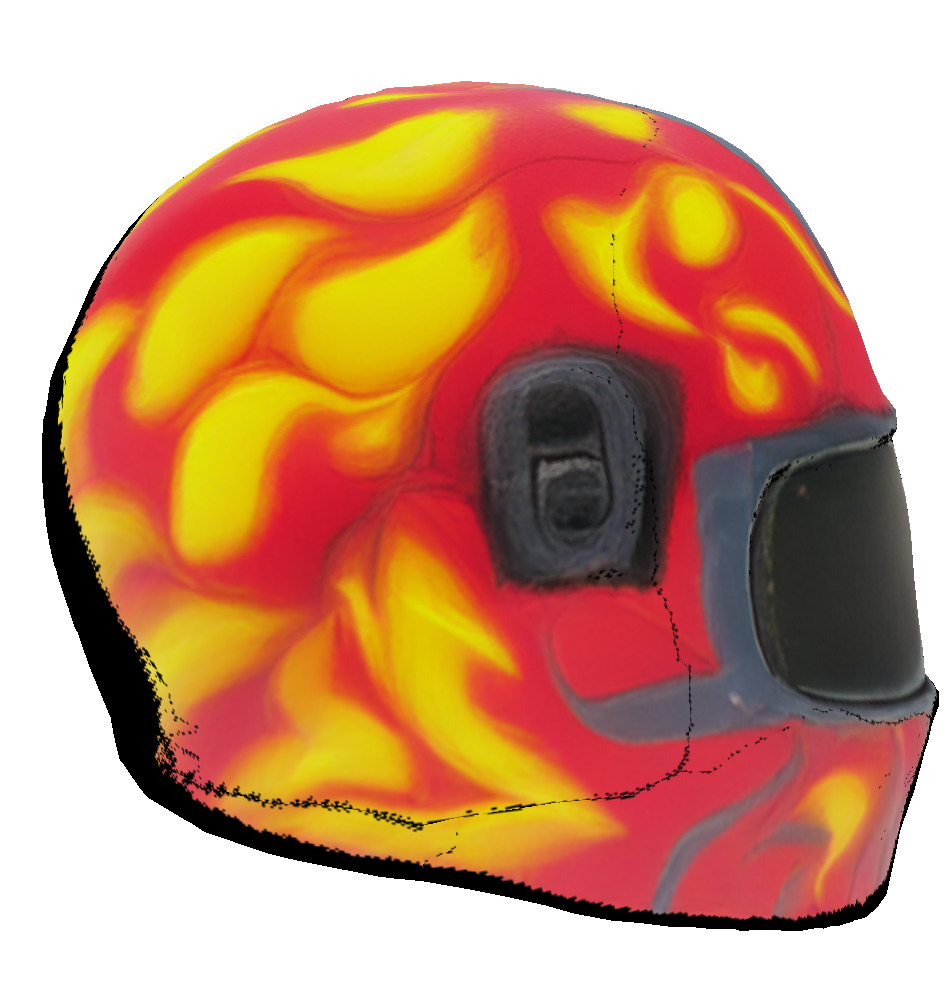}
        \vspace{-.1in}
        \caption*{raw back-projected texture images}
    \end{subfigure}
    \begin{subfigure}[b]{\columnwidth}
        \centering
        \includegraphics[width=0.35\textwidth]{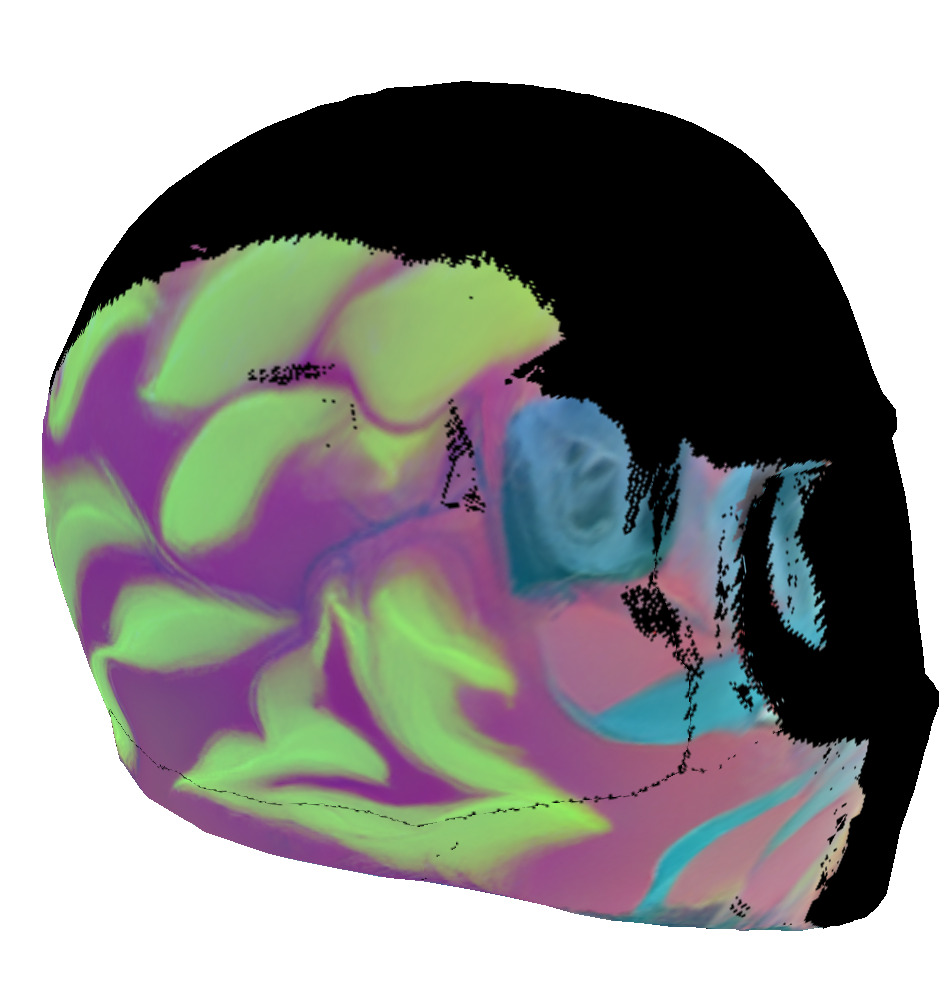}
        \hspace{0.05\textwidth}
        \includegraphics[width=0.35\textwidth]{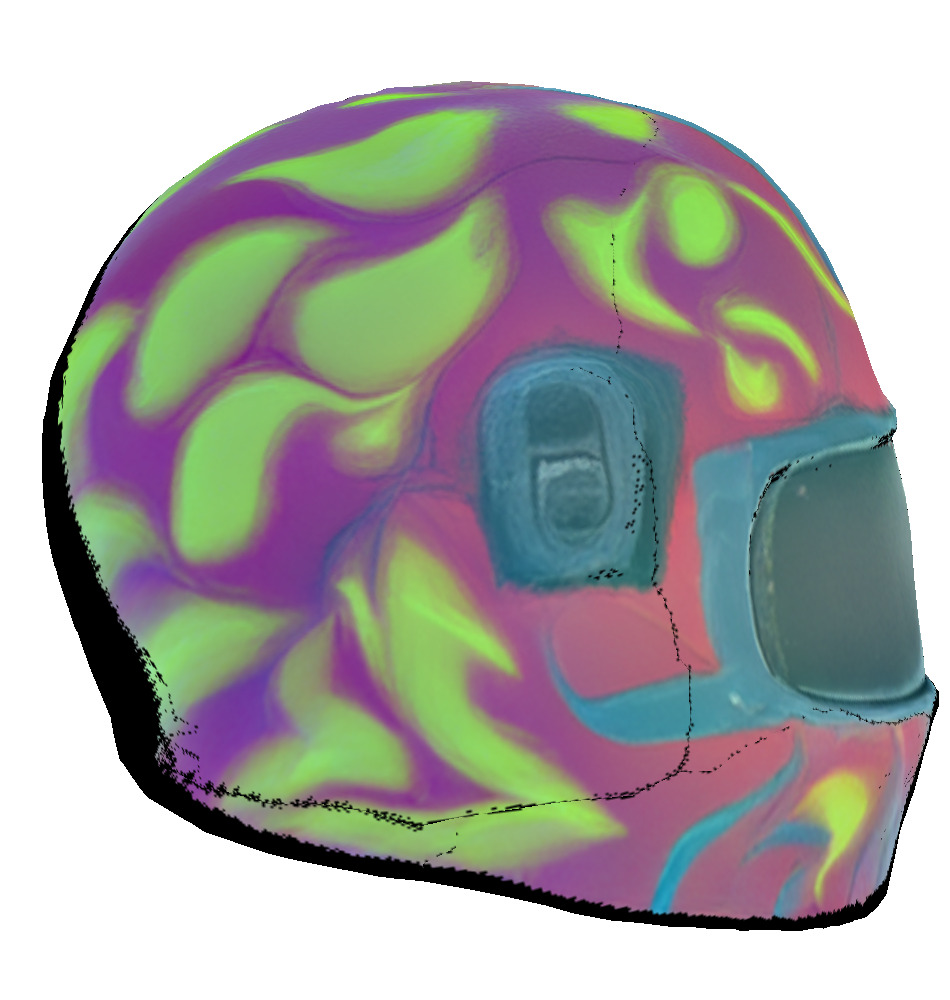}
        \vspace{-.1in}
        \caption*{normalized back-projected texture images}
    \end{subfigure}
    \caption{Back-projected renderings before (top) and after (bottom) normalization: The raw back-projections exhibit variation in luminance and saturation. Normalization reduces the discrepancy.} 
    \Tighten{-0.1in}
    \label{fig:mean and std of nets}
\end{figure}

\begin{comment}
\begin{figure}
    \centering
    
    \begin{subfigure}[b]{0.2\textwidth}
        \centering
        \includegraphics[width=\textwidth]{figures/normalization/t1.jpg}
    \label{fig:mean and std of net14}
    \end{subfigure}
    \hspace{.1in}    
    \begin{subfigure}[b]{0.2\textwidth}  
        \centering 
        \includegraphics[width=\textwidth]{figures/normalization/t2.jpg}
        \label{fig:mean and std of net24}
    \end{subfigure}
    \vspace{.1in}
    \begin{subfigure}[b]{0.2\textwidth}   
        \centering 
        \includegraphics[width=\textwidth]{figures/normalization/n1.jpg}
    \end{subfigure}
    \hspace{.1in}
    \begin{subfigure}[b]{0.2\textwidth}   
        \centering 
        \includegraphics[width=\textwidth]{figures/normalization/n2.jpg}
    \end{subfigure}
    \caption{Two back-projected renderings before (top) and after (bottom) color normalization. The input exhibits luminance and saturation variance. Normalization reduce the color inconsistency.} 
    \label{fig:mean and std of nets}
\end{figure}
\end{comment}

Normalization is crucial for the stability of optical flow. \figureref{fig:mean and std of nets} shows an example of two un-normalized back-projected renderings (top) and the corresponding renderings after normalization  (bottom). Due to color variations, the region of overlap is darker for the first rendering than it is for the second, resulting in $(\phi_t-\phi_s)$ being set to the wrong target difference in \equationref{eq:of_energy}. With normalization, the colors are more consistent.

\subsection{Registration}
Given back-projected color and confidence textures, $\big\{\{\NormalizedColor_i,\Confidence_i\}\big\}_{i=1}^K$, our goal is to compute the optical flow-fields $\{\FlowField_i\}_{i=1}^K$ that best align the colors. Here we focus on a single time-step of the flow computation. The complete hierarchical solve is described in \algoref{alg:process}.

    \paragraph*{Direct Approach}
A direct way to do this would be to define an energy that simultaneously optimizes \textit{all} the flow-fields. This could be done for a pair of textures by noting that the optimization for the source flow-field $\FlowField$ in \equationref{eq:of_energy} can be adapted to simultaneously optimize for the source and target flow-fields $\FlowField_s$ and $\FlowField_t$ by replacing the integrand accounting for advection along one field:
$$\Big(\big(f_s-\langle\FlowField,\nabla f_s\rangle\big)-f_t\Big)^2$$
with an integrand accounting for the advection along both:
$$\Big(\big(f_s-\langle\FlowField_s,\nabla f_s\rangle\big)-\big(f_t-\langle\FlowField_t,\nabla f_t\rangle\big)\Big)^2.$$

One could then obtain an energy simultaneously optimizing all the flow-fields by taking the sum of the energies over all pairs of textures. Such an energy would remain sparse and quadratic. However, the system complexity would be proportional to the mesh resolution \textit{multiplied by the number of input views}, which would be too expensive to solve in practice.

\begin{comment}
A direct way to do this would be to define an energy that simultaneously optimizes \textit{all} the optical flow-fields. This could be done by noting that for two color signals, $\Color_s$ and $\Color_t$, with associated confidences, $\Confidence_s$ and $\Confidence_t$, one can formulate the \textit{pairwise optical flow energy} is:
\begin{align*}
&E_\text{of}^{\{\Confidence_s,\Confidence_t\}}\left(\{\FlowField_s,\FlowField_t\}\,\big|\,\{f_s,f_t\},\alpha\right)\\
&\qquad\qquad\equiv
\SurfaceIntegral \Confidence_s\Confidence_t\Big(\big(f_s-\langle\FlowField_s,\nabla f_s\rangle\big)-\big(f_t-\langle\FlowField_t,\nabla f_t\rangle\big)\Big)^2\\
&\qquad\qquad+\SurfaceIntegral 
\alpha\left(\|\nabla\FlowField_s\|^2+\|\nabla\FlowField_t\|^2\right),
\end{align*}
with the joint confidence is set to the product of the confidences.
Then the optical flow energy simultaneously aligning all the colors signals is simply the sum of pairwise energies:
$$
E_\text{of}^{\{\Confidence_i\}}\Big(\{\FlowField_i\}\,\big|\,\{f_i\},\alpha\Big)\equiv
\sum_{i,j}E_\text{of}^{\{\Confidence_i,\Confidence_j\}}\Big(\{\FlowField_i,\FlowField_j\}\,\big|\,\{f_i,f_j\},\alpha\Big).$$

The energy is quadratic and results in a sparse symmetric positive-definite system for the coefficients of the flow-fields. However, using it requires solving a system whose complexity is proportional to the mesh resolution, multiplied by the number of color signals, which is too time-consuming in practice.
\end{comment}

    \paragraph*{Successive Signal Registration}
Instead, we propose an incremental approach. We start by selecting a target signal and successively register the remaining signals to it. The target itself is evolved by blending in the advected source colors and confidences.

\textit{Source/Target Selection.}
To select the initial target and successive sources, we require a measure of overlap between pairs of signals. Defining the \textit{coverage} of confidence $\Confidence:\Surface \rightarrow\R^{\geq0}$ as:
$$\langle\Confidence\rangle \equiv\SurfaceIntegral\Confidence$$
we define the \textit{overlap} of confidences $\Confidence$ and $\Confidence'$ to be the coverage of their product, $\langle\Confidence\cdot \Confidence'\rangle$. We then select the initial target by finding the pair of textures maximizing overlap, and use the one with larger coverage. Similarly, we select subsequent sources by finding the one maximizing the overlap with the current target. 

\textit{Source Registration.}
Given source and target color texture images, $\NormalizedColor_s$ and $\NormalizedColor_t$, with confidence texture images, $\Confidence_s$ and $\Confidence_t$, we solve for the flow-field registering the source to the target:
$$\FlowField = \text{Flow}(\NormalizedColor_s,\NormalizedColor_t,\Confidence_s\cdot \Confidence_t,\FlowFieldSmoothingWeight).$$

We advect the source's texture color and confidence images along $\FlowField$ and merge them into the target. The new target color is obtained by confidence-weighted blending of the textures registered so far:
$$
\NormalizedColor_t\leftarrow\frac{\sum_{i\in\mathcal{R}}\hbox{Adv}_{\FlowField_i}(\NormalizedColor_i)\cdot \hbox{Adv}_{\FlowField_i}(\Confidence_i)}{\sum_{i\in\mathcal{R}}\hbox{Adv}_{\FlowField_i}(\Confidence_i)}
$$
where $\mathcal{R}$ is the set of indices of the target and the sources registered to it, and the new target confidence is obtained by summing and clamping:
$$\Confidence_t\leftarrow \min\big(1,\ \Confidence_t + \hbox{Adv}_{\FlowField}(\Confidence_s)\big).$$
Though the blended value is undefined where both the source and target have zero confidence, this does not cause numerical instability because the blended confidence will be zero there as well.

We note that while the target is not advected within a single pass of optical flow registration, the choice of selected target can change across the hierarchical passes. As a result it is possible that all back-projected textures are deformed at the end of the registration.

\subsection{Stitching}
The final step is stitching the advected texture color and confidence images into a seamless texture color image. We do this using the adaptation of classical gradient-domain stitching that supports confidences, presented in \equationref{eq:confidence-weighted-stitching}.

\newcommand{\resultSize}{0.14}
\newcommand{\resultImage}[2]{%
    \begin{subfigure}[b]{\resultSize\textwidth}
        \centering
        \includegraphics[width=\textwidth]{figures/result/#1/#2_crop.jpg}
    \end{subfigure}
}

\newcommand{\resultImageWithCaption}[3]{%
    \begin{subfigure}[b]{\resultSize\textwidth}
        \centering
        \includegraphics[width=\textwidth]{figures/result/#1/#2_crop.jpg}
        \caption*{#3}%
    \end{subfigure}
}

\newcommand{\resultRow}[1]{%
    \resultImage{#1}{input1}%
    \resultImage{#1}{input2}%
    \resultImage{#1}{input_zoom1}%
    \resultImage{#1}{output_zoom1}%
    \resultImage{#1}{input_zoom2}%
    \resultImage{#1}{output_zoom2}%
}

\newcommand{\resultRowNew}[1]{%
    \resultImage{#1}{before1}%
    \resultImage{#1}{before2}%
    \resultImage{#1}{before3}%
    \resultImage{#1}{after3}%
    \resultImage{#1}{before4}%
    \resultImage{#1}{after4}%
}

\newcommand{\resultRowWithCaption}[1]{%
    \resultImageWithCaption{#1}{input1}{front}%
    \resultImageWithCaption{#1}{input2}{back}%
    \resultImageWithCaption{#1}{input_zoom1}{blended}%
    \resultImageWithCaption{#1}{output_zoom1}{registered \& stitched}%
    \resultImageWithCaption{#1}{input_zoom2}{blended}%
    \resultImageWithCaption{#1}{output_zoom2}{registered \& stitched}%
}

\newcommand{\resultRowNewWithCaption}[1]{%
    \resultImageWithCaption{#1}{before1}{front}%
    \resultImageWithCaption{#1}{before2}{back}%
    \resultImageWithCaption{#1}{before3}{blended}%
    \resultImageWithCaption{#1}{after3}{registered \& stitched}%
    \resultImageWithCaption{#1}{before4}{blended}%
    \resultImageWithCaption{#1}{after4}{registered \& stitched}%
}

\begin{figure*}[htp]
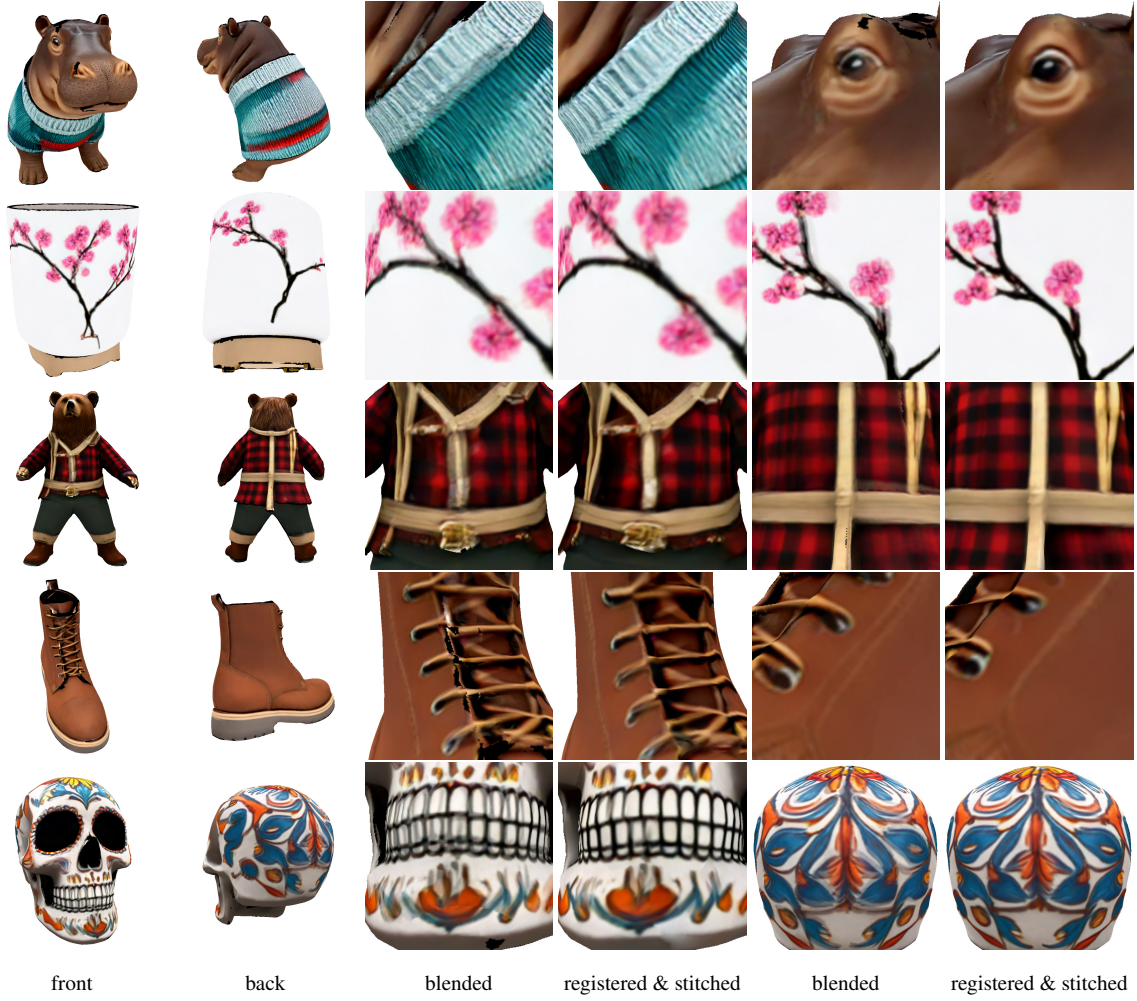

    \centering
    \resultRow{a4fa}
    \resultRow{dd03}
    \resultRow{b5d7}
    \resultRow{2e4f}%
    \resultRow{e88a}%

    \sneakInCaption{\resultSize}{front}%
    \sneakInCaption{\resultSize}{back}%
    \sneakInCaption{\resultSize}{blended}%
    \sneakInCaption{\resultSize}{registered \& stitched}%
    \sneakInCaption{\resultSize}{blended}%
    \sneakInCaption{\resultSize}{registered \& stitched}%
    
    \caption{Confidence-weighted blending vs. registration and stitching \revise{on the Instant3D dataset}: Front and back zoomed-out views of the confidence-weight blended back-projections are shown in the first two columns. Zoom-ins comparing blending to the proposed combination of registration and stitching are shown in the next four columns.}
    \label{fig:result}
\end{figure*}

\subsection{Process Summary}
Our processing pipeline is summarized in \algoref{alg:process}. Given the texture-atlased mesh, input renderings, and camera parameters, the renderings are back-projected onto the mesh and the colors are normalized (lines 1-5). Having initialized the flow-fields (lines 6 and 7) they are refined hierarchically (lines 8-27). The final flow-fields are then used to register the back-projected textures (lines 28 and 29), which are stitched together to generate the output (line 30).

\newcommand{\algindent}{\quad\quad}
\makeatletter
\algnewcommand{\StateNoNumber}[1]{\Statex \hskip\ALG@thistlm #1}
\makeatother
\RestyleAlgo{boxruled}
    \newcommand\mycommfont[1]{\fontsize{7.5}{8}\selectfont\ttfamily{#1}}
    \SetCommentSty{mycommfont}

\begin{algorithm}
    \caption{Renderings to Texture}
    \label{alg:process}
    \begin{algorithmic}[1]
        \Require 
        $\Surface,\,\,\big\{\{\Render_i,\Camera_i\}\big\}_{i=1}^K$
        \mbox{\hspace{-14pt}
        \line(1,0){3.05in}}
        \StateNoNumber{\tcp{Back-project and normalize colors}} 
        \State\textbf{for} $i\in\{1,K\}$:
        \State\algindent $\{\Color_i,\Confidence_i\}\leftarrow\hbox{ BackProjectAndGetConfidence}(\{\Render_i,\Camera_i\})$
        \State\algindent$\LocalColorMean_i\leftarrow\hbox{Blur}(\Color_i,\Confidence_i,\NormalizationWeight)$
        \State\algindent$\LocalColorScale_i\leftarrow\hbox{Blur}(\|\Color_i-\LocalColorMean_i\|,\Confidence_i,\NormalizationWeight)$
        \State\algindent$\NormalizedColor_i\leftarrow(\Color_i-\LocalColorMean_i)/\LocalColorScale_i$
        \StateNoNumber{\tcp{Initialize flow-fields}} 
        \State\textbf{for} $i\in\{1,K\}$:
        \State\algindent$\FlowField_i\leftarrow0$
        \StateNoNumber{\tcp{Hierarchical flow-field computation}}
        \State\textbf{for} $\ell\in\{1,L\}$:
        \StateNoNumber{\algindent\tcp{Advect normalized colors and confidences}}
        \State\algindent\textbf{for} $i\in\{1,K\}$:
        \State\algindent\algindent$\{\AdvectedNormalizedColor_i,\AdvectedConfidence_i\}\leftarrow\hbox{Adv}_{\FlowField_i}(\{\NormalizedColor_i,\Confidence_i\})$
        \StateNoNumber{\algindent\tcp{Set active indices and select target}}
        \State\algindent $\mathcal{I}\leftarrow\{1,\ldots,K\}$
        \State\algindent $\{i^*,j^*\}\leftarrow\argmax_{\{\{i,j\}\in\mathcal{I}\times\mathcal{I}|i<j\}}\langle \AdvectedConfidence_i\cdot \AdvectedConfidence_j\rangle$
        \State\algindent $t\leftarrow\argmax_{k\in\{i^*,j^*\}}\langle \AdvectedConfidence_k\rangle$
        \State\algindent$\mathcal{I}\leftarrow\mathcal{I}\backslash t$
        \State\algindent$\mathcal{R}\leftarrow\{t\}$
        \StateNoNumber{\algindent\tcp{Register to target}}
        \State\algindent\textbf{while} $\mathcal{I}\neq\emptyset$:
        \StateNoNumber{\algindent\algindent\tcp{Select next source and update indices}}
        \State\algindent\algindent$s\leftarrow\argmax_{i\in\mathcal{I}}\langle \AdvectedConfidence_i\cdot \AdvectedConfidence_t\rangle$
        \State\algindent\algindent$\mathcal{I}\leftarrow\mathcal{I}\backslash s$
        \State\algindent\algindent$\mathcal{R}\leftarrow\mathcal{R}\cup\{s\}$
        \StateNoNumber{\algindent\algindent\tcp{Blur and get incremental flow}}
        \State\algindent\algindent$\SmoothedAdvectedNormalizedColor_s\leftarrow\text{Blur}(\AdvectedNormalizedColor_s,\AdvectedConfidence_s,\ScalarSmoothingWeight)$
        \State\algindent\algindent$\SmoothedAdvectedNormalizedColor_t\leftarrow\text{Blur}(\AdvectedNormalizedColor_t,\AdvectedConfidence_t,\ScalarSmoothingWeight)$
        \State\algindent\algindent$\vec{\delta}\leftarrow \text{Flow}(\SmoothedAdvectedNormalizedColor_s,\SmoothedAdvectedNormalizedColor_t,\AdvectedConfidence_s\cdot\AdvectedConfidence_t,\FlowFieldSmoothingWeight)$
        \StateNoNumber{\algindent\algindent\tcp{Update source flow-field}}
        \State\algindent\algindent$\FlowField_s\leftarrow\FlowField_s+\vec{\delta}$
        \StateNoNumber{\algindent\algindent\tcp{Advect and merge normalized source}}
        \State\algindent\algindent$\{\AdvectedNormalizedColor_s,\AdvectedConfidence_s\}\leftarrow\hbox{Adv}_{\FlowField_s}(\{\NormalizedColor_s,\Confidence_s\})$
        \State\algindent\algindent$\AdvectedNormalizedColor_t\leftarrow \sum_{i\in\mathcal{R}}\hbox{Adv}_{\FlowField_i}(\NormalizedColor_i)\cdot\hbox{Adv}_{\FlowField_i}(\Confidence_i) \,\big/ \sum_{i\in\mathcal{R}}\hbox{Adv}_{\FlowField_i}(\Confidence_i)$
        \State\algindent\algindent$\AdvectedConfidence_t\leftarrow \min\big(1,\ \sum_{i\in\mathcal{R}}\hbox{Adv}_{\FlowField_i}(\Confidence_i)\big)$
        \StateNoNumber{\algindent\tcp{Reduce smoothness regularization}}
        \State\algindent $\ScalarSmoothingWeight\leftarrow\ScalarSmoothingWeight/2$,\algindent$\FlowFieldSmoothingWeight\leftarrow\FlowFieldSmoothingWeight/2$
        \StateNoNumber{\tcp{Align back-projected textures and stitch}}
        \State\textbf{for} $i\in\{1,K\}$:
        \State\algindent$\{\AdvectedColor_i,\AdvectedConfidence_i\}\leftarrow\hbox{Adv}_{\FlowField_i}(\{\Color_i,\Confidence_i\})$
        \State$\OutputColor\leftarrow\text{Stitch}\left(\big\{\{\AdvectedColor_i,\AdvectedConfidence_i\}\big\},\GradientFittingWeight,\GradientDomainRegularizationWeight\right)$
        \State\textbf{return} $\OutputColor$
    \end{algorithmic}
\end{algorithm}

The aim of each hierarchical iteration is to adjust the previously estimated flow-fields. As such, the iterations begin by advecting textures along the previously estimated flow-fields (lines 9 and 10). Then, a target signal maximizing overlap is selected (lines 12-13) and the remaining textures are successively integrated into the target, queued by overlap with the target (lines 16-26). To track the sources, a queue of texture indices is initialized (line 11) and elements are removed as their associated textures are processed (lines 14 and 18), while the merged textures are collected in a second set (lines 15 and 19). Finally, the weights affecting the smoothness of the correcting flow, $\ScalarSmoothingWeight$ and $\FlowFieldSmoothingWeight$, are reduced so that the next iteration can register finer detail (line 27).

To register, the advected source and target colors are smoothed (lines 20 and 21) and a correcting flow-field is computed (line 22). The  source's flow-field is adjusted to incorporate the correcting flow (line 23) and used to advect the source color and confidence (line 24). These are merged into the target (lines 25 and 26). 

\paragraph*{Function Spaces}
In generating the final texture image, our aim is to reproduce fine detail while keeping the dimensionality of linear systems small where possible. To this end we use two discretizations of scalar functions -- as per-vertex values at the mesh-resolution using the ``hat'' basis, and as per-texel values at the texture-image resolution using the basis of \citeauth{Prada}{prada2018}. For tangent vector-fields, we use the mesh-resolution basis of  \citeauth{Liu}{liu2026}.
We perform one-to-four planar subdivision to get meshes with \textasciitilde20K vertices. Using $1024\times1024$ textures, there are \textasciitilde700K active texels.

\textit{Colors and Confidences.}
We represent colors ($\Color_i$, $\NormalizedColor_i$, $\AdvectedNormalizedColor_i$, and $\AdvectedColor_i$) as well as confidences -- $\Confidence_i$ and $\AdvectedConfidence_i$ -- at texel-resolution, with the dimensions of the texture grid chosen based on the number of input renderings and their resolution.

\textit{Flow-fields.}
We represent the optical flow-fields ($\FlowField_i$ and $\vec{\delta}$) and the smoothed advected source and target colors ($\SmoothedAdvectedNormalizedColor_s$ and $\SmoothedAdvectedNormalizedColor_t$) at the mesh resolution. %

\paragraph*{Interaction between discretizations}
Interplay between different representations happens at several steps in the computation.

\textit{Normalization.}
To perform the normalization we compute the local mean and scale ($\LocalColorMean_i$ and $\LocalColorScale_i$) at mesh-resolution as we expect these to vary smoothly. We then evaluate them at texel positions to obtain the normalized texel-resolution colors ($\NormalizedColor_i$).

\textit{Flow-field computation.}
The smoothed colors ($\SmoothedAdvectedNormalizedColor_s$ and $\SmoothedAdvectedNormalizedColor_t$) used to compute the flow-field are represented at mesh-resolution. %

\begin{table*}[htp]
    \centering
    \begin{tabular}{@{}ccccc@{}}
        \toprule
         & View Selection & Camera Optimization & Registration & Blending \\
        \midrule
        \MVStex{} & yes & no & none &  gradient-domain (seam-localized)\\
        \ColorMap{} & yes & yes & rendering-space, non-rigid & visibility based average \\
        \AdvInitAlign{} & yes & no & rendering-space, translation & machine learning \\
        \revise{\ImSurfTex{}} & \revise{machine learning} & no & no & \revise{machine learning (localized window)}\\
        Ours & no & no & surface-domain, non-rigid & gradient-domain (global)\\ \bottomrule
    \end{tabular}
    \caption{Overview of the components used by different baseline methods.}
    \label{tab:baselines}
\end{table*}

\textit{Signal advection.}
Though the flow-field ($\FlowField_s$) is represented at mesh-resolution, it is applied to the texels, generating advected colors and confidences ($\AdvectedColor_s$ and $\AdvectedConfidence_s$) represented at texel-resolution.

In computing the system matrix and constraint coefficients, confidence is \textit{always} sampled at the texel-resolution to preserve detail. And, with the exception of computing the flow-fields, the construction of constraint coefficients samples color information at the texel-resolution as well.

\paragraph*{Quadrature}
For the optical flow systems, which are discretized at mesh-resolution, we use $Q=12$ quadrature points. For the stitching system, which is discretized at texel-resolution, we use $Q=3$.

\section{Evaluation}
\label{sec:evaluation}
We compare our optical flow registration to existing methods, focusing primarily on generated multi-view datasets.

\paragraph*{Data}
\revise{We evaluate our method on two generated datasets produced by different generation pipelines, each comprising ten objects, as well as on two RGB-D scans commonly used in prior work.}

\revise{For the first dataset (\emph{Instant3D}), we generate multi-view images and reconstruct the surface from them.} First, we generate nine renderings from a text prompt following the sparse multi-view generator in Instant3D \cite{li2024instantd}. The nine views are picked to span three elevation angles with uniformly sampled azimuths. Then we feed the renderings into MeshLRM \cite{wei2025meshlrm} to obtain the triangulated surface.

\revise{For the second dataset (\emph{SyncDreamer}), we pair generated multi-view images with an existing ground-truth surface. We take a ground-truth textured mesh from GSO~\cite{downs2022google} and render a single image at $30^\circ$ elevation. Conditioned only on this rendering, SyncDreamer~\cite{liu2024syncdreamer} generates sixteen uniformly distributed views at the same elevation. Because the views are generated without knowledge of the geometry, the renderings are less consistent with the surface than in the first dataset.}

\begin{figure}[htp]
    \centering
    \begin{subfigure}{0.165\textwidth}
        \includegraphics[width=\textwidth]{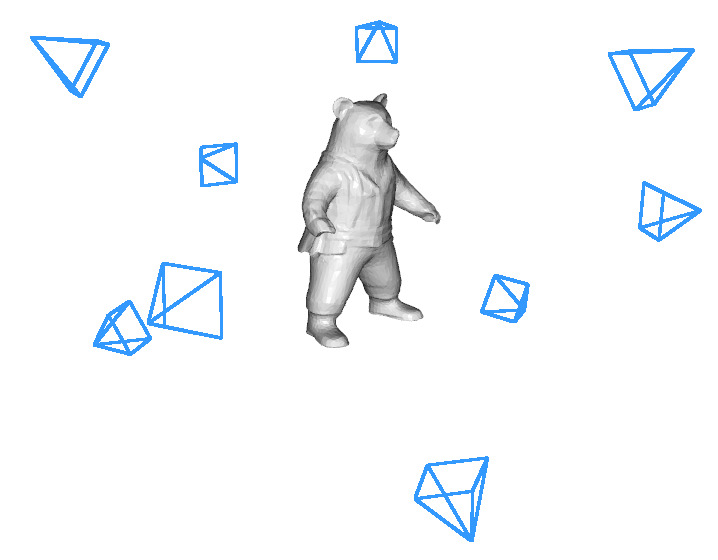}
    \end{subfigure}
    \hspace{.1\columnwidth}
    \begin{subfigure}{0.165\textwidth}
        \includegraphics[width=\textwidth]{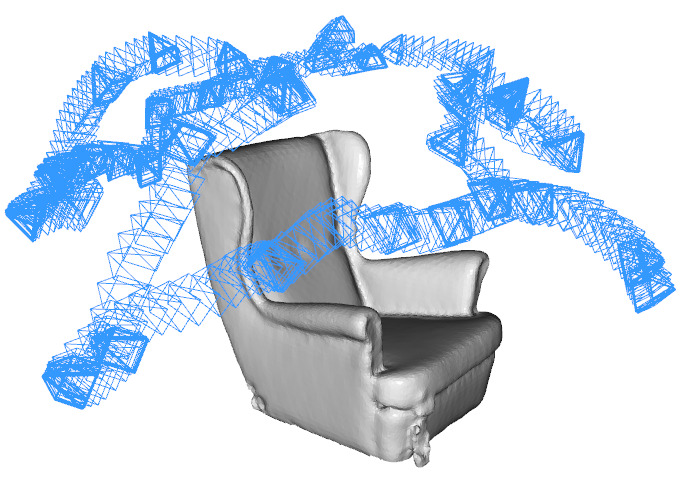}
    \end{subfigure}
    \caption{Visualization of the nine uniform camera views for \revise{Instant3D} generated dataset (left) and the original 550 camera views for the captured RGB-D data (right).} 
    \label{fig:camera_views}
\end{figure}

While we focus on the consolidation of generated renderings, we also consider real-world captures using the RGB-D scans of the Fountain \cite{CMO} and Armchair \cite{huang2020adversarial}. We use standard techniques to extract a textured mesh from the RGB-D data (see \appendixref{supp:RGBD} for details). We use all 33 keyframes from the Fountain dataset and randomly subsample 55 views from the full set of 550 scans of the Armchair. \figureref{fig:camera_views} shows the views for a generated dataset and the camera trajectory around the Armchair. %

\paragraph*{Baselines}
As Georgiou~\etal \cite{Im2SurfTex} note, many current texturing pipelines for generated renderings assign the most front-facing view to each primitive \cite{paint3d, chen2023text2tex, TEXTure} or apply naive blending \cite{MatAtlas, Texpainter, cheng2024mvpaint}, producing blurry or inconsistent textures \cite{MatAtlas}. We include naive weighted blending as a representative baseline, and also compare to Large-Scale Texturing (\MVStex{}) \cite{mvs-tex}, Color Map Optimization (\ColorMap{}) \cite{CMO}, Initialization and Alignment of Adversarial Texture (\AdvInitAlign{}) \cite{adv_align} \revise{and the Neural Backprojection (\ImSurfTex{}) \cite{Im2SurfTex}}.

Components used by the methods are summarized in \tableref{tab:baselines}. All baseline methods use view selection. \MVStex{} and \AdvInitAlign{} both assign an optimal view to each face. \ColorMap{} discards blurry views. \MVStex{} lacks an explicit registration stage while both \ColorMap{} and \AdvInitAlign{} incorporate rendering-space registration. \ColorMap{} optimizes for non-rigid distortion while \AdvInitAlign{} is restricted to translational alignment. Our method performs registration in the surface domain. 
 
All methods perform blending. \MVStex{} uses global color adjustment and seam-local gradient-domain stitching. \ColorMap{} uses visibility-based averaging. \AdvInitAlign{} uses adversarial texture optimization. We use global gradient-domain stitching.

\begin{comment}
    \begin{table*}[htp]
        \small
        \begin{tabular}{@{}cccccc@{}}
        \toprule
         & View Selection & Camera pose Optimization & Rendering-space Registration & Surface-domain registration & Blending \\ \midrule
        \MVStex{} & \checkmark &  &  &  & \checkmark \\
        \ColorMap{} & \checkmark & \checkmark & \checkmark &  & \checkmark \\
        \AdvInitAlign{} & \checkmark &  & \checkmark &  & \checkmark \\
        Ours &  &  &  & \checkmark & \checkmark \\ \bottomrule
        \end{tabular}
        \caption{Overview of different processing modules employed by different methods.}
        \label{tab:baselines}
    \end{table*}
\end{comment}

\paragraph*{Parameters}
For our approach, parameters are set based on texture and mesh resolutions. For baseline methods we use official implementations with parameters tuned for fair comparison. (See \appendixref{supp:parameters} for details). 

\begin{comment}
\subsubsection*{Metrics}
We adopt Learned Perceptual Image Patch Similarity (LPIPS) \cite{zhang2018perceptual} as a comparison metric. For each output texture, we render the model from the different input views and compute the average LPIPS score between input and rendered views. 

Ideally, we would use a leave-one-out approach, comparing renderings of the output to input that was omitted from the optimization. However, our focus on sparse input makes this impractical as fewer views introduce the additional challenge on coverage. As such, though we provide quantitative LPIPS measures, we rely primarily on qualitative assessments in our evaluation.
\end{comment}

\subsection{Our Method: Generated Data}

As discussed in \sectionref{sec:overview} and visualized in \figureref{fig:pipeline_comparison}, it is only with the combination of optical flow and gradient-domain stitching that we successfully produce seamless and ghost-free images. Additional results, shown in \figureref{fig:result}, confirm that our approach successfully aligns photometric features and removes discontinuities. (The figure also highlights an advantage of using gradient-domain stitching. In the zoom-in on the blended results for the Hippo there is a black region above the eye corresponding to geometry not imaged by any camera. With gradient-domain stitching, colors are filled in using an as-smooth-as-possible interpolant.)

\subsection{Comparisons: Sparse-view Generated Data}
\newcommand{\baselineCompareWidth}{0.13}
\setlength{\fboxsep}{0pt}
\setlength{\fboxrule}{0.1pt}
\newcommand{\baselineCompareUnit}[6]{
    \begin{minipage}{\baselineCompareWidth\textwidth}
        \centering
        \begin{subfigure}[b]{\linewidth}
            \centering
            \includegraphics[width=\linewidth]{figures/baseline_comparison_new/#1/#2_#3.jpg}
        \end{subfigure}%

        \begin{minipage}{\linewidth}
            \begin{subfigure}[b]{0.485\linewidth}
                \centering
                \fbox{{\includegraphics[width=\linewidth]{figures/baseline_comparison_new/#1/#2_#4.jpg}}}
            \end{subfigure}
            \begin{subfigure}[b]{0.485\linewidth}
                \centering
                \fbox{{\includegraphics[width=\linewidth]{figures/baseline_comparison_new/#1/#2_#5.jpg}}}
            \end{subfigure}
        \end{minipage}%
        
    \end{minipage}%
}

\begin{figure*}[htb]
    \centering

    \begin{minipage}{\textwidth}
        \centering
        \makebox[15pt][c]{\rotatebox[origin=c]{90}{\sffamily\small Mask}}
        \baselineCompareUnit{78df}{input}{box_box_crop}{crop}{box_crop}{0.1}
        \baselineCompareUnit{78df}{mvs}{box_box_crop}{crop}{box_crop}{0.1}
        \baselineCompareUnit{78df}{cmo}{box_box_crop}{crop}{box_crop}{0.1}
        \baselineCompareUnit{78df}{adv}{box_box_crop}{crop}{box_crop}{0.1}
        \baselineCompareUnit{78df}{im2surftex}{box_box_crop}{crop}{box_crop}{0.1}
        \baselineCompareUnit{78df}{nerf}{box_box_crop}{crop}{box_crop}{0.1}
        \baselineCompareUnit{78df}{ours}{box_box_crop}{crop}{box_crop}{0.1}
    \end{minipage}%
    
    \vspace{1em}
    \begin{minipage}{\textwidth}
        \centering
        \makebox[15pt][c]{\rotatebox[origin=c]{90}{\sffamily\small Penguin}}
        \baselineCompareUnit{5487}{input}{box_box_crop}{crop}{box_crop}{0.1}
        \baselineCompareUnit{5487}{mvs}{box_box_crop}{crop}{box_crop}{0.1}
        \baselineCompareUnit{5487}{cmo}{box_box_crop}{crop}{box_crop}{0.1}
        \baselineCompareUnit{5487}{adv}{box_box_crop}{crop}{box_crop}{0.1}
        \baselineCompareUnit{5487}{im2surftex}{box_box_crop}{crop}{box_crop}{0.1}
        \baselineCompareUnit{5487}{nerf}{box_box_crop}{crop}{box_crop}{0.1}
        \baselineCompareUnit{5487}{ours}{box_box_crop}{crop}{box_crop}{0.1}
    \end{minipage}%
    
    \vspace{1em}
    \begin{minipage}{\textwidth}
        \centering
        \makebox[15pt][c]{\rotatebox[origin=c]{90}{\sffamily\small Pharaoh}}
        \baselineCompareUnit{3c48}{input}{box_box_crop}{crop}{box_crop}{0.1}
        \baselineCompareUnit{3c48}{mvs}{box_box_crop}{crop}{box_crop}{0.1}
        \baselineCompareUnit{3c48}{cmo}{box_box_crop}{crop}{box_crop}{0.1}
        \baselineCompareUnit{3c48}{adv}{box_box_crop}{crop}{box_crop}{0.1}
        \baselineCompareUnit{3c48}{im2surftex}{box_box_crop}{crop}{box_crop}{0.1}
        \baselineCompareUnit{3c48}{nerf}{box_box_crop}{crop}{box_crop}{0.1}
        \baselineCompareUnit{3c48}{ours}{box_box_crop}{crop}{box_crop}{0.1}
    \end{minipage}%

    \begin{minipage}{\textwidth}
        \centering
        \hspace{17pt}
        \sneakInCaption{\baselineCompareWidth}{Blended}
        \sneakInCaption{\baselineCompareWidth}{MVS-tex}
        \sneakInCaption{\baselineCompareWidth}{ColorMap}
        \sneakInCaption{\baselineCompareWidth}{\AdvInitAlign{}}
        \sneakInCaption{\baselineCompareWidth}{\ImSurfTex{}}
        \sneakInCaption{\baselineCompareWidth}{NeRF (MeshLRM)}
        \sneakInCaption{\baselineCompareWidth}{Ours}%
    \end{minipage}%

    \caption{Comparisons on sparse generated data \revise{(Instant3D)}: MVS-tex exhibits visible inconsistency at patch boundaries. ColorMap and \AdvInitAlign{} degrade on sparse multi-view data. \revise{\ImSurfTex{}} and MeshLRM renderings exhibit ghosting at in-between views. Our method resolves misalignment and seamlessly merges the back-projections. (See the supplementary video for additional views.)}
    \label{fig:baseline_comparison}
\end{figure*}
\newlength{\LOOw}
\setlength{\tabcolsep}{1pt}
\setlength{\LOOw}{\dimexpr(\textwidth - 12\tabcolsep)/8\relax}

\newcommand{\LOOimg}[2]{\includegraphics[width=\LOOw]{figures/syncdreamer_LOO/figures/#1_#2.jpg}}

\newcommand{\LOOrow}[1]{%
  \LOOimg{#1}{GT}   & 
  \LOOimg{#1}{leaved_out} & 
  \LOOimg{#1}{blended} &
  \LOOimg{#1}{mvs}  & 
  \LOOimg{#1}{cmo}  & 
  \LOOimg{#1}{adv}  &
  \LOOimg{#1}{im2surftex}  &
  \LOOimg{#1}{ours} \\
}

\newcommand{\LOOheader}{%
  Reference & Held-out view & Blended & \MVStex{} & \ColorMap{} & \AdvInitAlign{} & \ImSurfTex{} & Ours \\
}

\begin{figure*}[t]
  \centering
  \setlength{\tabcolsep}{1pt}
  \renewcommand{\arraystretch}{0.4}
  \begin{tabular}{@{}cccccccc@{}}
    \LOOrow{alarm}
    \LOOrow{dogbowl}
    \LOOrow{mario}
    \LOOrow{sneaker}
    \LOOheader
  \end{tabular}
  \caption{\revise{Leave-four-out comparison on the SyncDreamer dataset. Each object is registered from 12 of its 16 views and rendered from the 4 withheld views. "Reference" shows the original textured mesh with one withheld view. "Held-out view" shows the withheld generated SyncDreamer rendering from the same view. The remaining columns show the registration results of different methods.}}
  \label{fig:syncdreamer_LOO}
\end{figure*}

We compare to the baseline methods in \figureref{fig:baseline_comparison} %
and \figureref{fig:syncdreamer_LOO}. The rows alternate between showing zoomed-out views, and showing zoom-ins on regions where the input is misaligned. \MVStex{} generates consistent texture within patches imaged from a single view, but produces seams at patch boundaries. Though \ColorMap{} and \AdvInitAlign{} perform registration, both are designed for dense RGB-D video streams and performance degrades on sparse, generated, multi-view data. \revise{\ImSurfTex{} produces blurry textures and ghosting artifacts as it only aggregates colors from small fixed pixel-radius neighborhoods in image space, so that misalignments larger than the neighborhood cannot be reconciled.} Additionally, we observe that our method reproduces colors more faithfully -- \MVStex{}'s output tends to have higher contrast, \AdvInitAlign{} introduces chromatic shifts (e.g., for the Mask example, \AdvInitAlign{}'s colors become greener), \revise{and \ImSurfTex{} can produce darker renderings (see the
Pharaoh example)}.

\begin{table*}[]
    \setlength{\tabcolsep}{3pt}
    \centering
    \footnotesize
 \begin{tabular}{lccccccccc}
    \toprule
    Dataset
    & \multicolumn{3}{c}{Instant3D}
    & \multicolumn{3}{c}{SyncDreamer}
    & \multicolumn{3}{c}{SyncDreamer-L4O} \\
    \cmidrule(lr){2-4} \cmidrule(lr){5-7} \cmidrule(lr){8-10}
    Metric & LPIPS$\downarrow$ & SSIM$\uparrow$ & ECD$\downarrow$ & LPIPS$\downarrow$ & SSIM$\uparrow$ & ECD$\downarrow$ & LPIPS$\downarrow$ & SSIM$\uparrow$ & ECD$\downarrow$ \\ \midrule
    blended
      & \cellcolor{second}0.192 & \cellcolor{best}0.684 & \cellcolor{second}6.36
      & 0.249 & 0.619 & 1.43
      & 0.298 & 0.477 & 2.10 \\
    \MVStex{}
      & 0.195 & 0.592 & 8.20
      & 0.213 & 0.514 & 1.68
      & 0.325 & 0.364 & 2.26 \\
    \ColorMap{}
      & 0.223 & 0.562 & 6.99
      & 0.200 & 0.582 & 1.40
      & \cellcolor{second}0.260 & 0.492 & 2.07 \\
    \AdvInitAlign{}
      & 0.320 & 0.593 & 7.23
      & 0.315 & 0.484 & 1.69
      & 0.360 & 0.395 & 2.11 \\
    \ImSurfTex{}
      & 0.245 & \cellcolor{second}0.647 & 7.68
      & \cellcolor{second}0.199 & \cellcolor{best}0.668 & \cellcolor{second}1.32
      & 0.268 & \cellcolor{best}0.519 & \cellcolor{second}2.05 \\
    Ours
      & \cellcolor{best}0.167 & 0.615 & \cellcolor{best}5.95
      & \cellcolor{best}0.198 & \cellcolor{second}0.632 & \cellcolor{best}1.30
      & \cellcolor{best}0.256 & \cellcolor{second}0.508 & \cellcolor{best}1.98 \\
    \bottomrule
    \end{tabular}
    \caption{\revise{LPIPS, SSIM, and ECD scores on the generated datasets, averaged over all objects and views. Best and second-best results per column are highlighted in dark and light purple, respectively. (Lower LPIPS and ECD, and higher SSIM, are better.)}}
    \label{tab:quantitative-comparison}
\end{table*}

\revise
{
We perform an empirical comparison using Learned Perceptual Image Patch Similarity (LPIPS)~\cite{zhang2018perceptual}, Structural Similarity (SSIM)~\cite{SSIM}, and Extended Convolution Descriptor (ECD)~\cite{Mitchel:2020:ESA,ECDMatching:2026}.
}
\revise{For each method, we render the output textured mesh from the perspective of the input cameras, and compare it to the input renderings. For SyncDreamer we additionally evaluate in a leave-four-out (L4O) setting. Of the 16 generated views we withhold four at $90^\circ$ intervals around the orbit, and register the texture from the remaining 12. Each method's texture is rendered at the four withheld views and compared against a render of the original textured mesh from the same view. We do not apply this protocol to Instant3D, as its nine views do not provide sufficient coverage when a view is withheld.} 

\revise{\tableref{tab:quantitative-comparison} gives the scores averaged over all objects and views, with the best and second-best results highlighted. Our approach ranks best on LPIPS on every dataset. SSIM has been observed to favor averaging-based reconstructions~\cite{adv_align}, which favors both naive blending and the learned aggregation of \ImSurfTex{}, and does not fully reflect the perceptual quality differences visible in \figureref{fig:baseline_comparison}.} (See \appendixref{app:quantitative} for more results \revise{and discussion of the metrics}.)

Though we focus on view-independent texture generation, \figureref{fig:baseline_comparison} also shows renderings from the NeRF model produced by MeshLRM \cite{wei2025meshlrm}. Using a view-dependent representation of the surface color, NeRF reproduces the input renderings. However, for the frontal views shown in the figure, which are not in the set of input views, the renderings exhibit undesired artifacts.  (For example, the stripes of the flag on the Penguin are misaligned  and the ridges on the head-piece of the Pharaoh exhibit ghosting.)

\newcommand{\rgbdCompareWidth}{0.15}

\newcommand{\rgbdCompareUnit}[5]{
    \begin{minipage}{\rgbdCompareWidth\textwidth}
        \begin{subfigure}[b]{\linewidth}
            \centering
            \includegraphics[width=\linewidth]{figures/RGBD_comparison/#1/#2_#3.jpg}
        \end{subfigure}%
    
        \begin{subfigure}[b]{0.485\linewidth}
            \centering
            \resizebox{\linewidth}{!}{\includegraphics[width=\linewidth]{figures/RGBD_comparison/#1/#2_#4.jpg}}
        \end{subfigure}%
        \vspace{0.01\linewidth}
        \begin{subfigure}[b]{0.485\linewidth}
            \centering
            \resizebox{\linewidth}{!}{\includegraphics[width=\linewidth]{figures/RGBD_comparison/#1/#2_#5.jpg}}
        \end{subfigure}%
    \end{minipage}%
}

\newcommand{\rgbdCompareRow}[4]{%
    \rgbdCompareUnit{#1}{average}{#2}{#3}{#4}
    \rgbdCompareUnit{#1}{mvs}{#2}{#3}{#4}
    \rgbdCompareUnit{#1}{cmo}{#2}{#3}{#4}
    \rgbdCompareUnit{#1}{adv}{#2}{#3}{#4}
    \rgbdCompareUnit{#1}{im2surftex}{#2}{#3}{#4}
    \rgbdCompareUnit{#1}{ours}{#2}{#3}{#4}%
}

\begin{figure*}[htp]
    \centering
    \begin{minipage}{\textwidth}
        \centering
        \makebox[15pt][c]{\rotatebox[origin=c]{90}{\sffamily\small Fountain}}
        \rgbdCompareRow{fountain}{crop_box_box}{crop_crop}{crop_box_crop}
    \end{minipage}%
    \vspace{0.3em}
    
    \begin{minipage}{\textwidth}
        \centering
        \makebox[15pt][c]{\rotatebox[origin=c]{90}{\sffamily\small Armchair}}
        \rgbdCompareRow{chair}{crop_box_box}{crop_crop}{crop_box_crop}
    \end{minipage}%
    
    \begin{minipage}{\textwidth}
        \centering
        \hspace{17pt}
        \sneakInCaption{\rgbdCompareWidth}{blended}
        \sneakInCaption{\rgbdCompareWidth}{MVS-tex}
        \sneakInCaption{\rgbdCompareWidth}{ColorMap}
        \sneakInCaption{\rgbdCompareWidth}{\AdvInitAlign{}}
        \sneakInCaption{\rgbdCompareWidth}{\ImSurfTex{}}
        \sneakInCaption{\rgbdCompareWidth}{ours}%
    \end{minipage}%

    \caption{Comparisons on real-world RGB-D scans: MVS-Tex generates sharp results through view selection but has seams (Fountain, both zoom-ins) and exhibits ghosting (Armchair, right zoom-in). Though the number of views is increased, \AdvInitAlign{} continues to produce speckled results \revise{and \ImSurfTex{} produces blurry textures}. In contrast, ColorMap, as well as our method, generate results without seams or ghosting. For the more densely sampled data, ColorMap's result are sharper than ours.}
    \label{fig:RGBD_comparison}
\end{figure*}

\subsection{Comparison: Dense-view RGB-D Data}
Comparisons on real-world RGB-D scans are shown in \figureref{fig:RGBD_comparison}. \MVStex{} produces sharp results through view selection (Armchair, left zoom-in) but exhibits seams (Fountain, left zoom-in) and ghosting (Fountain, right zoom-in and Armchair, right zoom-in). \AdvInitAlign{} consistently produces speckling (Armchair and Fountain, both zoom-ins). \ColorMap{} and our method produce results
\begin{wrapfigure}[5]{r}{0.41\columnwidth}
    \hspace{-0.05\columnwidth}
    \begin{minipage}{1.1\linewidth}
    \hfill
    \includegraphics[width=0.48\linewidth]{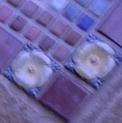}
    \includegraphics[width=0.48\linewidth]{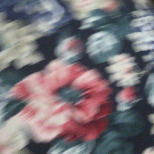}
    \end{minipage}
\end{wrapfigure}%
free of registration artifacts. While the baselines avoid integrating
blurry images through view selection, our method does not. Still, it remains robust to motion-blur in the input (see inset).

\subsection{Robustness to Different Types of Distortions}

\begin{figure}[htbp]
  \centering
  \captionsetup[subfigure]{font=small,labelformat=empty,skip=2pt}
  \begin{subfigure}[b]{0.207\linewidth}
    \centering
    \includegraphics[width=\linewidth]{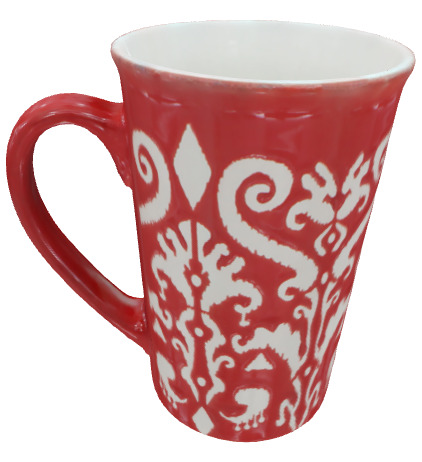}
    \caption{Ground Truth}
    \label{fig:degrade-perfect}
  \end{subfigure}\hfill
  \begin{subfigure}[b]{0.348\linewidth}
    \centering
    \includegraphics[width=\linewidth]{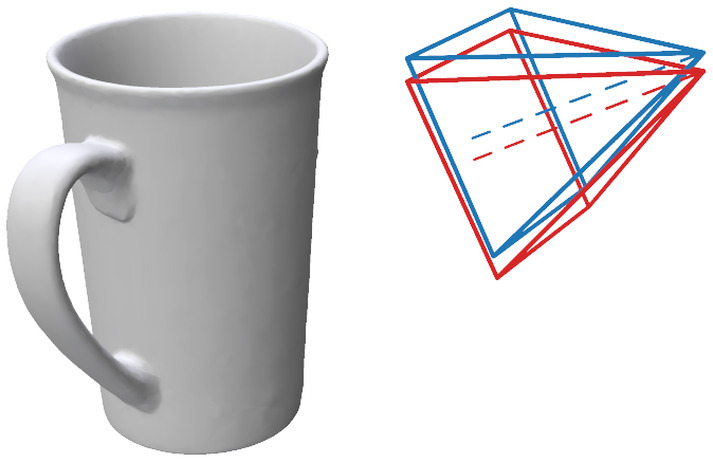}
    \caption{Camera}
    \label{fig:degrade-camera}
  \end{subfigure}\hfill
  \begin{subfigure}[b]{0.207\linewidth}
    \centering
    \includegraphics[width=\linewidth]{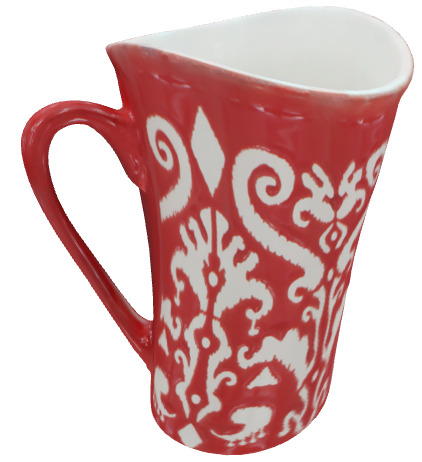}
    \caption{Mesh}
    \label{fig:degrade-mesh}
  \end{subfigure}\hfill
  \begin{subfigure}[b]{0.207\linewidth}
    \centering
    \includegraphics[width=\linewidth]{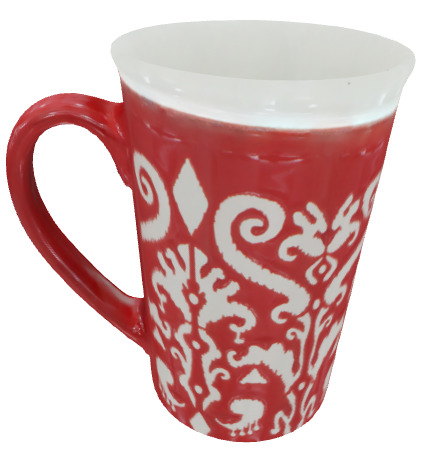}
    \caption{Texture}
    \label{fig:degrade-texture}
  \end{subfigure}
  \caption{\revise{Simulated distortions used in the robustness study; the camera is distorted (red) by applying a random translation and rotation to the true camera pose (blue); the mesh is distorted  by advecting the vertices along a spatial flow-field; and the texture is distorted by advecting along a random surface flow-field.}}
  \label{fig:degradations}
\end{figure}

\begin{figure*}[htbp]
  \centering
  \includegraphics[width=\linewidth]{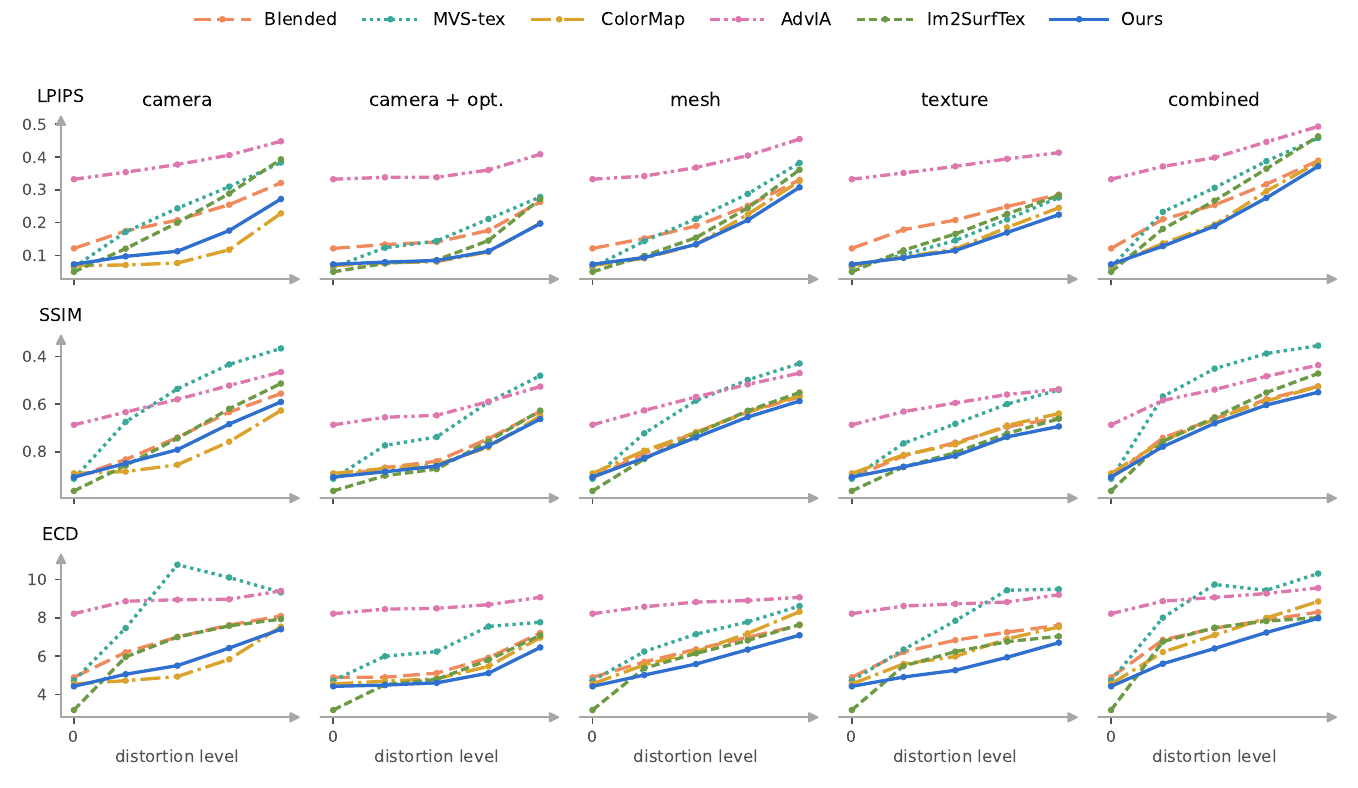}
  \caption{\revise{Robustness to distortion, plotting results for different distortions (columns) and similarity measures (rows): Results are averaged over the GSO textured meshes with distortion severity increasing along the $x$-axis.
  The plots show results without (camera) and with (camera + opt.) performing ColorMap's camera pose refinement after the initial camera distortion. For all metrics, lower is better (the SSIM axis is flipped).
  }}
  \label{fig:robustness-curves}
\end{figure*}

\revise{To evaluate robustness, we introduce four types of simulated distortions to the input data: surface-domain texture distortion; per-vertex mesh distortion; camera pose distortion; and all three combined. Using the same GSO textured meshes from the SyncDreamer experiments, we apply each distortion at increasing severity levels and compare the resulting textures against the ground truth using the same L4O setting discussed above. Figure~\ref{fig:degradations} visualizes the effect of each type of distortion.  Since real capture pipelines typically refine camera poses before texturing, we additionally evaluate a camera-corrected variant of the camera distortion, in which the distorted poses are first refined by the rigid stage of ColorMap Optimization and all methods are then run on the refined poses.}

\revise{Figure~\ref{fig:robustness-curves} plots each metric as a function of distortion level. Most methods perform comparably on clean data, but as distortion severity grows, our method emerges as the best-performing on texture, mesh, and combined distortion, demonstrating the strongest robustness to input inconsistencies. Regarding camera pose distortion, ColorMap Optimization leads at every distortion level, which is expected since it is the only method that directly optimizes camera poses. Every method (except ColorMap) improves once it is given the refined poses (Figure~\ref{fig:robustness-curves}, camera + opt.). With refined poses, ours remains the most robust when the pose error is the largest.}

\subsection{Performance}

\tableref{tab:performance} compares running times for three generated and two real-world datasets. Our performance is comparable to that of state-of-the-art methods on sparse input, despite the complexity of repeatedly solving the optical flow problem. However, as our method does not leverage view selection, \textit{all} back-projections are considered when computing the value of an output texel, and our performance deteriorates with increased input size. All results are obtained using a desktop with an Intel Core i7-8700K CPU and NVIDIA GeForce RTX 2080 GPU.

\begin{table}[]
    \centering
    \small
    \begin{tabular}{@{}cccccc@{}}
    \toprule
    & Instant3D & SyncDreamer & Fountain & Armchair \\ \midrule
    Views & 9 & 16 & 33 & 55 \\ \midrule
    \MVStex{} & 11 & 14 & 12 & 13 \\
    \ColorMap{} & 46 & 90 & 348 & 408 \\
    \AdvInitAlign{} & 429 & 522 & 510 & 727 \\
    \ImSurfTex{} & 50 & 119 & 424 & 780 \\
    Ours & 71 & 83 & 842 & 1725 \\ \bottomrule
    \end{tabular}
    \caption{ \revise{Runtime (in seconds) for the each methods. For the generated datasets (Instant3D and SyncDreamer), we report the
    average runtime over the ten objects. For the RGB-D scans, we report the runtime on each scan. Runtimes for \ImSurfTex{} include the precomputation of the geodesic information.} }
    \label{tab:performance}
\end{table}
    
\section{Discussion, Limitations, and Future Work}  

\paragraph*{Repeating Patterns}
\newcommand{\stripeImage}[3][]{%
    \begin{subfigure}[b]{#3\linewidth}
        \centering
        \includegraphics[width=\linewidth]{figures/stripes/#2.jpg}
        \ifstrempty{#1}{}{%
            \caption*{\small #1}%
        }%
    \end{subfigure}
}

\begin{figure}[htp]
    \centering
    \begin{subfigure}[t]{\columnwidth}
        \centering
        \stripeImage[view \#1: 6 white stripes]{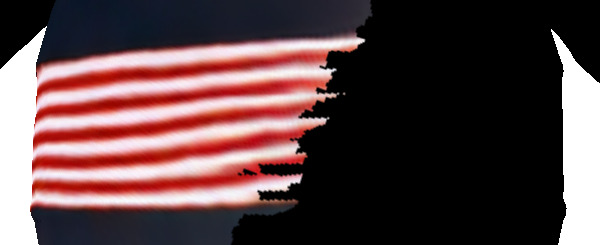}{0.48}
        \stripeImage[view \#2: 6 white stripes]{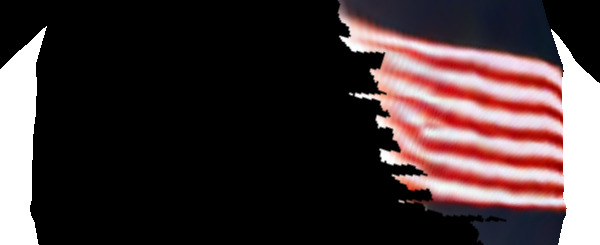}{0.48}

        \vspace{0.5em}
        \stripeImage[view \#3: 7 white stripes]{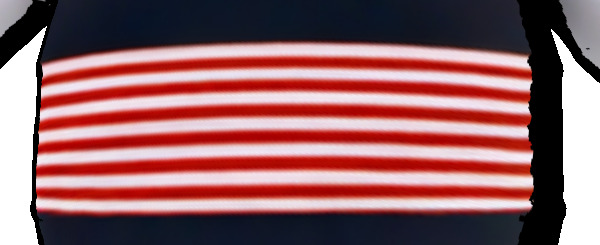}{0.48}
        \stripeImage[view \#4: 7 \& 6 white stripes]{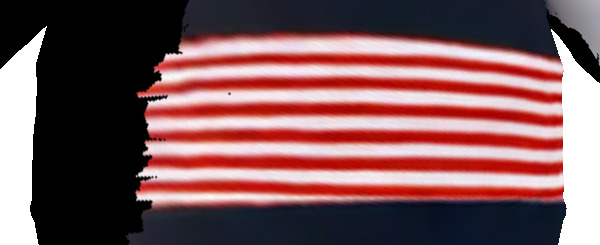}{0.48}

        \vspace{0.5em}
        \stripeImage[blended]{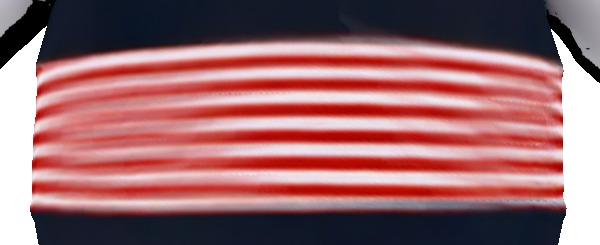}{0.48}
        \stripeImage[registered \& stitched]{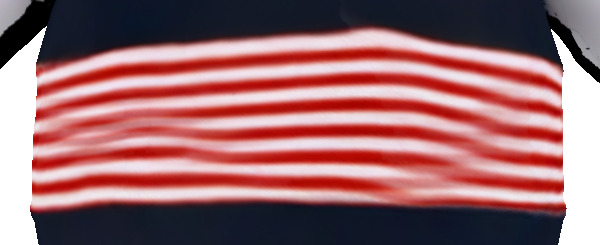}{0.48}
        \caption{mismatched stripe counts}
        \label{fig:stripes_a}
    \end{subfigure}

    \vspace{1em}
    
    \begin{subfigure}[t]{\columnwidth}
        \centering
        \stripeImage[view \#1]{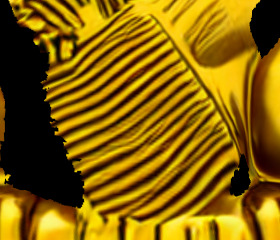}{0.3}
        \stripeImage[view \#2]{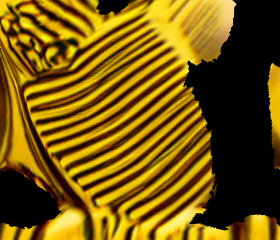}{0.3}
        \stripeImage[view \#3]{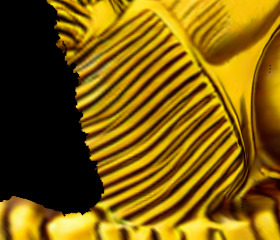}{0.3}

        \vspace{0.5em}
        \stripeImage[blended]{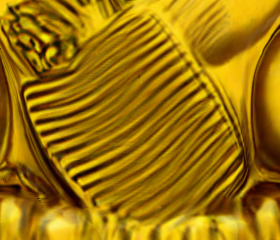}{0.3}
        \stripeImage[registered \& stitched]{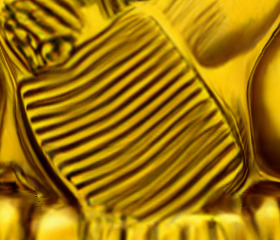}{0.3}
        \caption{inconsistent stripe layout}
        \label{fig:stripes_b}
    \end{subfigure}
        
    \caption{Stripe registration: When the number of stripes varies or their layout is inconsistent, some twisting and blurring can occur.}
    \label{fig:stripes}
\end{figure}

In the context of generated images, repeating patterns are a particular challenge as different renderings are not guaranteed to produce the same number of elements or have the same layouts. We demonstrate two such difficult cases in \figureref{fig:stripes}. 

For the back of the Penguin shown in \figureref{fig:baseline_comparison}, renderings \#1 and \#2 have different numbers of stripes than renderings \#3 and \#4, making perfect registration impossible. This results in both twisting (due to optical flow) and blurring (due to stitching). Notably, on the front of the Penguin, where the number of stripes is consistent across the views, our approach aligns the stripes and is not trapped in the local minima that tend to arise when registering textures containing repeating patterns.

For the front of the Maneki-neko shown in \figureref{fig:stripes_b}, the renderings have different layouts of stripes. Still, our method produces an output in which the individual stripes are sharp, with only a little localized blurring resulting from merging inconsistent layouts.
    
\paragraph*{Compatibility}
\newcommand{\combineImage}[3][]{%
    \begin{subfigure}[b]{0.11\textwidth}
        \centering
        \includegraphics[width=\textwidth]{figures/camera_pose/#2_#3.jpg}
        \ifstrempty{#1}{}{
            \caption*{\small #1}%
        }%
    \end{subfigure}%
}

\begin{figure}[t]
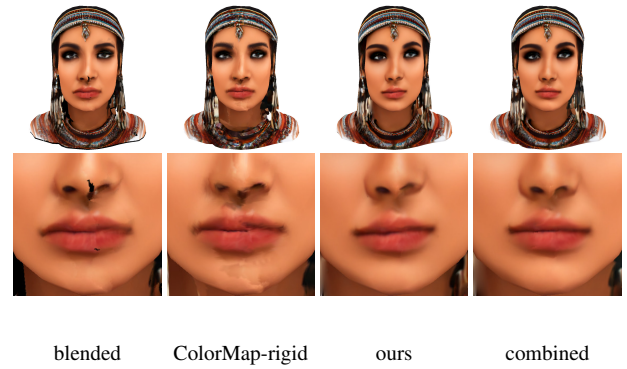

    \centering
    \combineImage{unregistered}{cropped}
    \combineImage{CMO_cam}{cropped}
    \combineImage{ours}{cropped}
    \combineImage{ours+cam}{cropped}
    \combineImage[blended]{unregistered}{cropped_crop}
    \combineImage[ColorMap-rigid]{CMO_cam}{cropped_crop}
    \combineImage[ours]{ours}{cropped_crop}
    \combineImage[combined]{ours+cam}{cropped_crop}
    
    \caption{Compatibility with existing methods: Combining our methods with ColorMap's rigid camera pose optimization produces a texture image with the facial symmetry of ColorMap and the seamlessness of our method.}
\label{fig:combination}
\end{figure}

Our surface-domain optical flow and gradient-domain stitching are easy to integrate within existing approaches. For example, \figureref{fig:combination} shows that combining our registration and stitching with \ColorMap{}'s camera pose optimization produces more symmetric textures than using either method alone.

\paragraph*{Failure Cases}
\newcommand{\coverageUnit}{
    \begin{minipage}{0.24\textwidth}
        \begin{subfigure}[b]{0.6\linewidth}
            \centering
            \includegraphics[width=\linewidth]{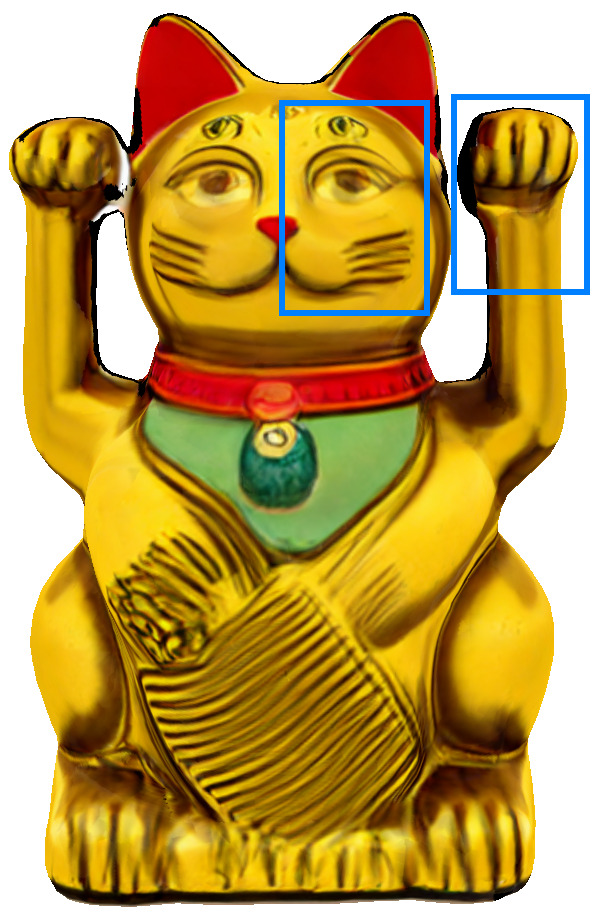}
        \end{subfigure}%
        \begin{minipage}{0.3\linewidth}
            \begin{subfigure}[b]{\linewidth}
                \centering
                \resizebox{\linewidth}{!}{\includegraphics[width=\linewidth]{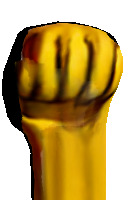}}
            \end{subfigure}
            
            \begin{subfigure}[b]{\linewidth}
                \centering
                \resizebox{\linewidth}{!}{\includegraphics[width=\linewidth]{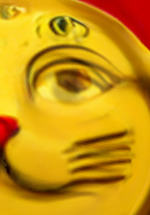}}
            \end{subfigure}%
        \end{minipage}
    \end{minipage}%
}

\begin{figure}[htp]
    \centering
    \begin{subfigure}[t]{\columnwidth}
        \centering
        \begin{subfigure}[t]{0.8\linewidth}
            \centering
            \begin{subfigure}[t]{0.183\textwidth}
                \includegraphics[width=\linewidth]{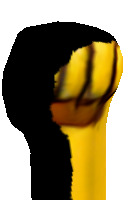}
                \caption*{view \#1}
            \end{subfigure} 
            \hfill
            \begin{subfigure}[t]{0.183\textwidth}
                \includegraphics[width=\linewidth]{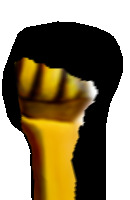}
                \caption*{view \#2}
            \end{subfigure}             
            \hfill
            \begin{subfigure}[t]{0.183\textwidth}
                \includegraphics[width=\linewidth]{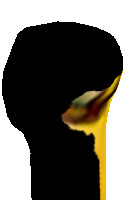}
                \caption*{view \#3}
            \end{subfigure}
            \hfill
            \begin{subfigure}[t]{0.183\textwidth}
                \includegraphics[width=\linewidth]{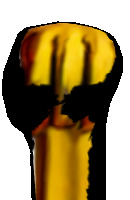}
                \caption*{view \#4}
            \end{subfigure}              
        \end{subfigure}

        \begin{subfigure}[t]{0.49\linewidth}
            \includegraphics[width=0.62\linewidth]{figures/failure_cases/a54c_new/input_crop_main_crop_box_box.jpg}
            \includegraphics[width=0.3\linewidth]{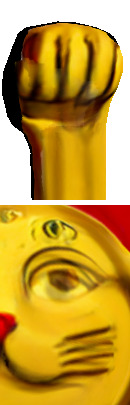}
            \caption*{blended}
        \end{subfigure}%
        \begin{subfigure}[t]{0.49\linewidth}
            \includegraphics[width=0.62\linewidth]{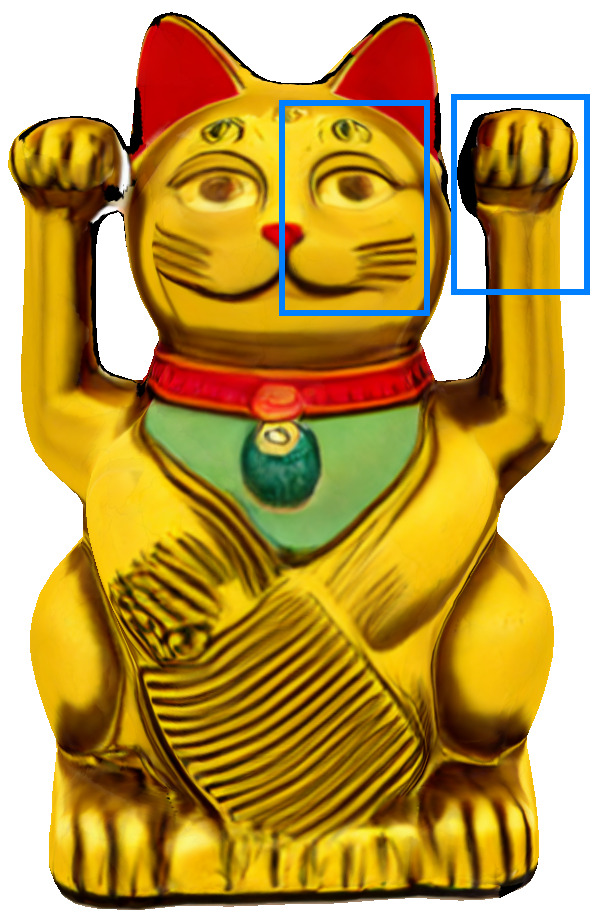}
            \includegraphics[width=0.3\linewidth]{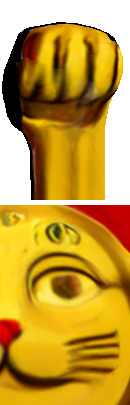}
            \caption*{registered \& stitched}
        \end{subfigure}
        
        \caption{Minimal coverage and lack of detail}
        \label{fig:failure_a}
    \end{subfigure}

    \begin{subfigure}[t]{\linewidth}
        \begin{subfigure}[t]{0.32\linewidth}
            \includegraphics[width=\linewidth]{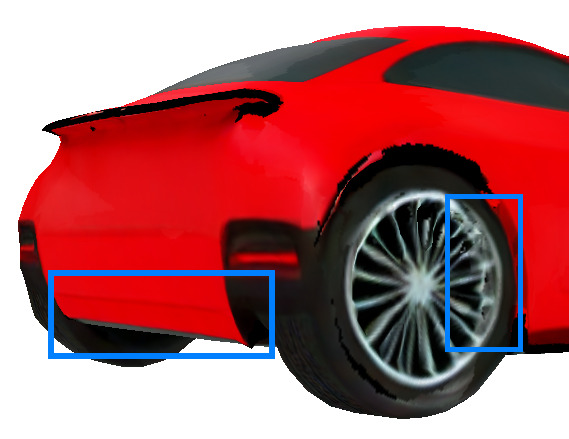}
            \caption*{blended}
        \end{subfigure}%
        \hfill
        \begin{subfigure}[t]{0.32\linewidth}
            \includegraphics[width=\linewidth]{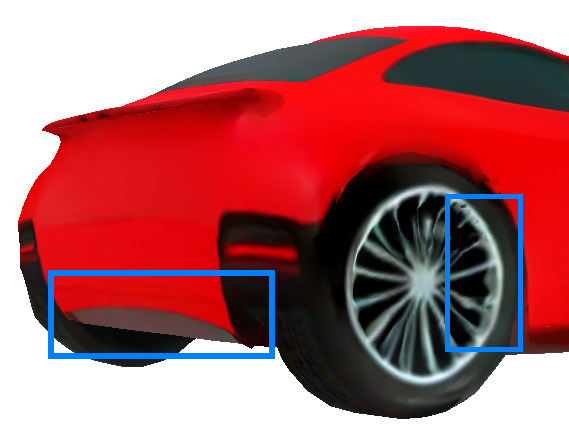}
            \caption*{output}
        \end{subfigure}%
        \hfill
        \begin{subfigure}[t]{0.32\linewidth}
            \includegraphics[width=\linewidth]{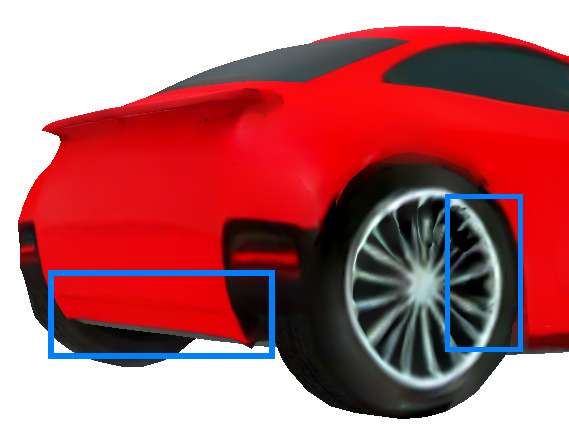}
            \caption*{+ stationary target}
        \end{subfigure}%
         \caption{Texture drift, mitigated by incorporating a stationary target}
        \label{fig:failure_b}
    \end{subfigure}
    \caption{Failure cases of our method: Insufficient view overlap causes registration degradation at the paw (top). Simultaneous flow optimization results in the undercarriage drifting out from the bottom of the car (bottom, left and center). This can be mitigated by using a coarse texture as a stationary target (bottom, right).} 
    \label{fig:failure}
\end{figure}

As with appearance-based registration methods in general, our method struggles when there is little overlap between the renderings or when there is insufficient photometric detail in the regions of overlap to guide the optical flow. This can be seen in \figureref{fig:failure_a} where the limited overlap combined with the lack of features results in ghosting around the paws. %

Additionally, the optical flow energy is (approximately) invariant to the application of the same deformation to all textures, with only the regularization weight introducing a bias towards smaller deformation. As a result, texture-registration may exhibit drifting, where colors are aligned but move jointly from their ``natural'' location. This can be seen in \figureref{fig:failure_b} where the ghosting at the wheel is resolved, but the undercarriage moves from under the car.

When it is possible to obtain a coarse color texture (e.g. by sampling a NeRF model), drifting can be mitigated by using the coarse colors as a \textit{stationary} target to which the textures are registered. This can be seen in \figureref{fig:failure_b} (right), where using such a target mitigates the problem of drift while still resolving the ghosting.

\begin{comment}
    \subsubsection*{Guiding the registration}
    
    \input{figures/NeRF_guided/NeRF_guided}
    
    It is possible to guide our registration process. For our generated dataset, each mesh is extracted from MeshLRM, which predicts a NeRF with learned geometric priors. We also extract a texture from this NeRF by sampling colors near the mesh surface and then use it to guide the registration: all views are registered to the extracted texture directly, with unseen areas filled with it during stitching.
    
    We demonstrate the results of the guided registration in \figureref{fig:guided}. The extracted texture is coherent and covers the entire mesh, but lacks the high-frequency detail present in direct multi-view registration. Our method removes most misalignment between views, but can cause subtle drift in texture position. Using the extracted texture as guidance mitigates this drift. During stitching, unseen areas are filled using harmonic interpolation without texture detail. Filling large holes with extracted texture provides more natural results.

\end{comment}

\paragraph*{Future Work}
A natural next-step is the incorporation of existing methods for quantifying blurriness within our confidence measure, particularly as this can reduce the running time when using denser multi-view datasets. We would also like to consolidate and simplify parameters, to make it easier to apply our method to data acquired with different modalities.

More generally, we would like to explore learning confidence values from the input renderings and their depth-maps. We would also like to explore augmenting the optical flow energy with a term measuring the alignment of photometric features to geometric features as an alternative means for mitigating drift.

\bibliographystyle{eg-alpha-doi} 
\bibliography{TROF}

\appendix

\section{RGB-D Scan Processing}
\label{supp:RGBD}
To obtain texture-atlased geometry for  the Fountain and Armchair, we perform the following steps: 
\begin{enumerate}
    \item We remove background geometry from the RGB-D scans.
    \item We reconstruct the surface \cite{kazhdan2013screened} and remove spurious surface components reconstructed in under-sampled regions.
    \item We perform isotropic remeshing \cite{10.1145/166117.166119} followed by quadratic edge collapse \cite{garland1997surface} to ensure that meshes are of similar quality and have comparable resolution.
    \item We texture-atlas the geometry \cite{sawhney2017boundary}.
    \end{enumerate} 
    Steps 1-3 were performed using MeshLab \cite{meshlab}. 
\section{Implementation Details}
\label{supp:parameters}
We discuss details of our implementation and parameter selection for results shown in \sectionref{sec:evaluation}.

\subsection{Back-Projection}
We compute the depth confidence for visible texels in the $i$-th view by measuring texels' distance from the nearest silhouette and having the confidence fall off for texels within a prescribed radius:
\begin{enumerate}
\item We compute the depth-map for each camera $\Camera_i$.
\item We compute the Laplacian of the depth-map and scale it to have standard deviation equal to one.
\item We detect discontinuity boundaries as locations where the Laplacian is greater than one.
\item For each texel, we compute its position on the surface, back-project it into the depth-map, and compute the distance to the nearest discontinuity, $D_\text{disc}$. We then set the confidence to $\min\left(1,\frac{D_\text{disc}}{r_\text{dc}}\right)$, fixing $r_\text{dc}$ at $20$ pixels.
\item We quantize values below $0.15$ to zero.
\end{enumerate}

\subsection{Optical Flow Registration}
\begin{wrapfigure}[9]{r}{0.40\columnwidth}
    \vspace{-1.3\baselineskip}
    \hspace{-0.05\columnwidth}
    \begin{minipage}{0.42\columnwidth}
    \includegraphics[width=.48\columnwidth]{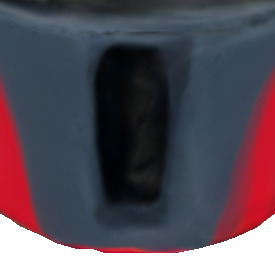}
    \includegraphics[width=.48\columnwidth]{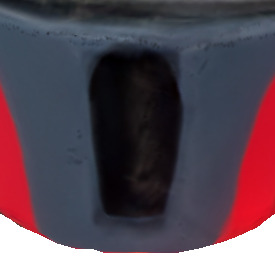}
    \includegraphics[width=.48\columnwidth]{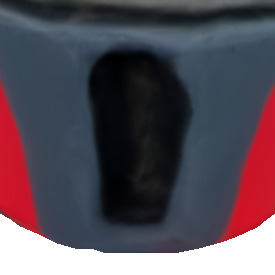}\hfill
    \includegraphics[width=.48\columnwidth]{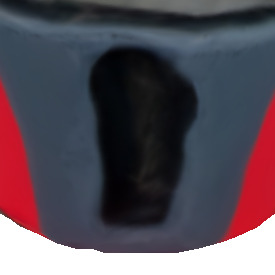}
    \end{minipage}
\end{wrapfigure}
There are four parameters for optical flow registration: the number of hierarchy levels, $L$; the weights affecting the smoothness of the estimated flow-field, $\ScalarSmoothingWeight$ and $\FlowFieldSmoothingWeight$; and the extent of the smoothing used to estimate local mean and scale, $\NormalizationWeight$. The number of levels is a trade-off, with too few resulting in an overly smooth flow-field that fails to register fine features, and too many resulting in a flow-field that better aligns the features at the cost  of allowing high-frequency oscillation in the deformation. Empirically, we found that setting $L=4$ and $\ScalarSmoothingWeight=0.002$ worked for most experiments. We set $\FlowFieldSmoothingWeight=0.2$ for generated data and $\FlowFieldSmoothingWeight=0.1$ for the RGB-D scans, and we set the normalization weight to $\NormalizationWeight=0.001$. (The inset shows registration results for the mouthpiece of the Helmet in \figureref{fig:pipeline_comparison}, computed using different numbers of hierarchy levels, $L\in\{2,4,6,8\}$.)

\subsection{Gradient-domain Stitching}
Though we had to fine-tune the gradient fitting weight in a small number of cases, we found that $\GradientFittingWeight=0.01$ works well in general. We fix the smoothness regularization weight at $\GradientDomainRegularizationWeight=10^{-8}$ so as to only take effect in areas of zero confidence. %

\subsection{Baselines}
In our evaluations, We use the publicly available implementations of the baselines \cite{mvs-tex,open3d,adv_align,Im2SurfTex}.

For \ColorMap{} we found that the alternating optimization did not produce optimal results on our synthesized data. Instead, we evaluated three variants (alternating optimization, rigid-only optimization, and rigid followed by non-rigid optimization) and present the best visual result. 

For \AdvInitAlign{} we lower the texture resolution to $256\times256$ for each patch when processing the synthesized data to reduce the speckling artifacts resulting from the sparse sampling of camera views. For the RGB-D scans we use a resolution of $1024\times1024$. 

\revise{For \ImSurfTex{}, we use the released implementation and pretrained weights, running
only the neural back-projection stage. We modify the implementation to accept full extrinsic camera parameters and to reduce its memory cost to fit our hardware. We set the texture resolution to $256\times256$ to reduce speckling artifacts on the synthesized datasets, and to $1024\times1024$ for the RGB-D scans. Geodesic distances are computed using the affine heat method~\cite{soliman2025affine}.}

As some baselines require a finer mesh resolution for better results, we tested all methods with zero, one, two, and three passes of subdivision and present the best results. For \MVStex{} and \ColorMap{}, meshes are subdivided using planar one-to-four subdivision. For \AdvInitAlign{}, we use the provided Loop subdivision.%
\section{Quantitative Comparisons}
\label{app:quantitative}

\begin{figure}[t]
  \centering
  \setlength{\tabcolsep}{2pt}
  \begin{tabular}{@{}ccc@{}}
    \includegraphics[width=0.31\linewidth]{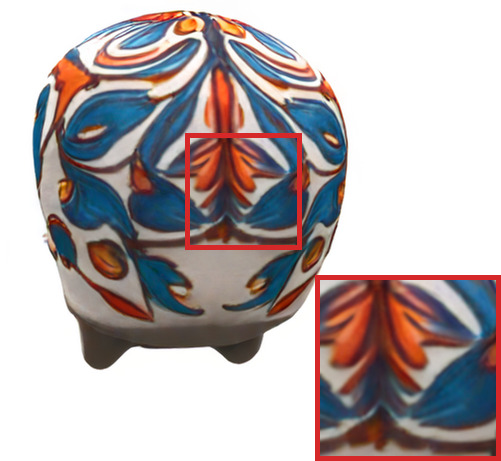} &
    \includegraphics[width=0.31\linewidth]{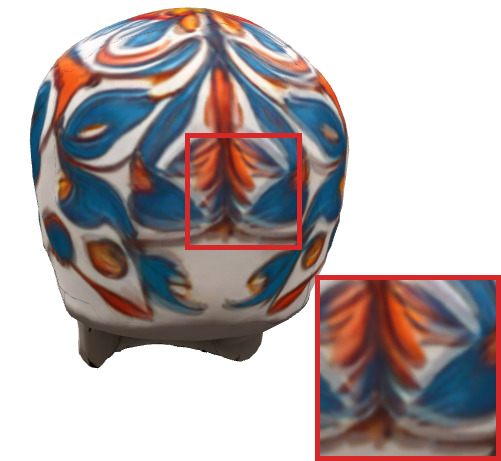} &
    \includegraphics[width=0.31\linewidth]{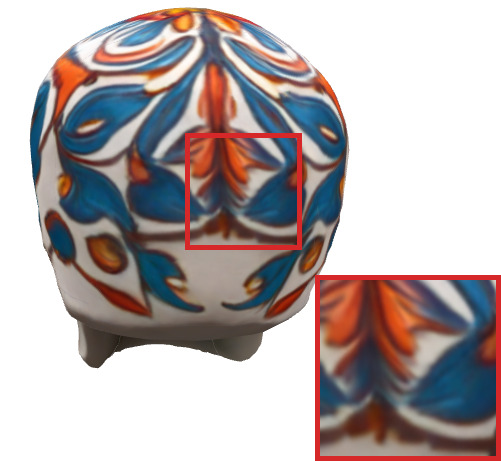} \\
    input & blended & ours \\
    & \textbf{0.145} / \textbf{0.690} / 6.30 & 0.208 / 0.325 / \textbf{5.52} \\
  \end{tabular}
  \caption{\revise{A limitation of image-space metrics for evaluating texture consistency: We compare the input rendering, the blended result, and ours on an Instant3D object,  with LPIPS$\downarrow$ / SSIM$\uparrow$ / ECD$\downarrow$ scores below each (better in bold). The blended rendering exhibits ghosting, yet performs better with respect to LPIPS and SSIM.}}
  \label{fig:ghosting-metric-failure}
\end{figure}

\revise{The classical LPIPS and SSIM metrics  do not reliably penalize ghosting artifacts. As such, we additionally leverage the ECD metric, comparing descriptors from the reference renderings to descriptors obtained by rendering the geometry using the synthesized textures. Because it is constructed using the local distribution of gradient orientations and magnitudes, ECD is more sensitive to blurring and ghosting. Figure~\ref{fig:ghosting-metric-failure} shows an example of this -- despite the ghosting, LPIPS and SSIM exhibit a preference for the blended result while ECD identifies the sharper texture produced using our approach as more similar to the reference.}

We additionally compute the absolute difference in sharpness score ($S_3$)~\cite{S3} %
comparing output textured mesh renderings to input renderings. However, \revise{the S3 score does not reflect} the cross-view texture consistency shown in the qualitative results of \sectionref{sec:evaluation}. In particular, the difference in $S_3$ favors \MVStex{} whose per-triangle view selection maintains input sharpness within each triangle by construction.

\begin{table}[H]
    \centering
    \begin{tabular}{@{}cccc@{}}
        \toprule
                    & Mask  & Penguin & Pharaoh \\ \midrule
        blended      & 0.141 & 0.105   & 0.131   \\
        \MVStex{}    & \textbf{0.075} & 0.091   & \textbf{0.088}   \\
        \ColorMap{}     & 0.204 & 0.085   & 0.096   \\
        \AdvInitAlign{} & 0.113 & 0.135   & 0.113   \\
        \ImSurfTex{} & 0.113 & \textbf{0.068} &  0.162 \\
        ours         & 0.105 & 0.074   & 0.131   \\ \bottomrule
    \end{tabular}
    \caption{Difference in $S_3$ scores comparing renderings of textured meshes to input renderings. Lower values indicate closer sharpness to the input.}
    \label{tab:empirical-comparison-S3}
\end{table}

\end{document}